\documentclass[11pt]{article}
\DeclareMathAlphabet{\scr}{U}{rsfs}{m}{n}

\pdfoutput=1
\usepackage{latexsym}
\usepackage{epsfig}
\usepackage[mathscr]{eucal}
\usepackage{amsfonts}
\usepackage{amscd}
\usepackage{cite}
\usepackage{array}
\usepackage{amssymb}
\usepackage{colordvi}
\usepackage[centertags]{amsmath}
\usepackage{enumerate}
\usepackage{graphicx}
\usepackage{booktabs}
\usepackage{theorem}
\usepackage[footnotesize]{caption}
\usepackage{soul}
\usepackage{url}
\usepackage[hidelinks]{hyperref}
\hypersetup{
  colorlinks   = true, 
  urlcolor     = blue, 
  linkcolor    = black, 
  citecolor   = black 
}

\usepackage[numbers,sort&compress]{natbib}
\let\OLDthebibliography\thebibliography
\renewcommand\thebibliography[1]{
  \OLDthebibliography{#1}
  \setlength{\parskip}{4pt}
  \setlength{\itemsep}{0pt plus 0.3ex}
}

\usepackage{slashed}
\usepackage{color}
\usepackage{bbm}
\usepackage[utf8]{inputenc}
\allowdisplaybreaks

\newcommand{\newc}{\newcommand}
\newc{\be}{\begin{equation}}
\newc{\ee}{\end{equation}}
\newc{\bea}{\begin{eqnarray}}
\newc{\eea}{\end{eqnarray}}
\newc{\ol}{\overline}
\newc{\wt}{\widetilde}
\newc{\bs}{\boldsymbol}
\newc{\m}{\mathcal}
\newc{\la}{\langle}
\newc{\ra}{\rangle}

\newcommand{\beq}{\begin{eqnarray}}
\newcommand{\eeq}{\end{eqnarray}}
\newcommand{\bpmatrix}{\begin{pmatrix}}
\newcommand{\epmatrix}{\end{pmatrix}}
\newcommand{\ba}{\begin{array}}
\newcommand{\ea}{\end{array}}

\renewcommand{\ol}{\text{1l}}

\renewcommand{\eqref}[1]{Eq.~(\ref{#1})}
\newcommand{\bib}[1]{Ref.~\cite{#1}}
\newcommand{\bibs}[1]{Refs.~\cite{#1}}

\newcommand{\bc}{\begin{center}}
\newcommand{\ec}{\end{center}}

\newcommand{\gev}{~\textrm{GeV}}

\newcommand\refeq[1]{Eq.~(\ref{#1})}

\newcommand\refta[1]{Tab.~\ref{#1}}
\newcommand\refse[1]{Sect.~\ref{#1}}

\newcommand\citere[1]{Ref.~\cite{#1}}
\newcommand\citeres[1]{Refs.~\cite{#1}}

\def\reffi#1{\mbox{Fig.~\ref{#1}}}

\begin{document}

\begin{flushright}
KA-TP-10-2023
\end{flushright}

\title{\textbf{
2HDM interpretations of the CMS diphoton\\[0.4em]
excess at 95 GeV}}

\date{}
\author{
Duarte Azevedo$^{1,2\,}$\footnote{E-mail:
\texttt{duarte.azevedo@kit.edu}} ,
Thomas Biekötter$^{1\,}$\footnote{E-mail:
\texttt{thomas.biekoetter@kit.edu}} ,
P.M. Ferreira$^{3,4\,}$\footnote{E-mail:
\texttt{pmmferreira@fc.ul.pt}} 
\\[5mm]
{\small\it
$^1$Institute for Theoretical Physics, Karlsruhe Institute of Technology,} \\
{\small\it Wolfgang-Gaede-Str.~1,
76128 Karlsruhe, Germany}\\[3mm]
{\small\it
$^2$Institute for Astroparticle Physics, Karlsruhe Institute of Technology,} \\
{\small\it Hermann-von-Helmholtz-Platz 1, 76344 Eggenstein-Leopoldshafen, Germany}\\[3mm]
{\small\it
$^3$Centro de F\'{\i}sica Te\'{o}rica e Computacional,
    Faculdade de Ci\^{e}ncias,} \\
{\small \it    Universidade de Lisboa, Campo Grande, Edif\'{\i}cio C8
  1749-016 Lisboa, Portugal} \\[3mm]
{\small\it
$^4$ISEL -
 Instituto Superior de Engenharia de Lisboa,} \\
{\small \it   Instituto Polit\'ecnico de Lisboa
 1959-007 Lisboa, Portugal} \\[3mm]
}

{
\let\newpage\relax
\let\clearpage\relax
\maketitle
}

\begin{abstract}
In both Run~1 and Run~2 of the LHC,
the CMS collaboration has observed an excess
of events in the searches for low-mass
Higgs bosons in the diphoton final state
at a mass of about 95~GeV.
After a recent update of the experimental
analysis, in which the full Run~2 data
collected at 13~TeV has been included
and an improved experimental calibration
has been applied, the local significance of
the excess amounts to $2.9\sigma$.
The presence of this diphoton excess is
especially interesting in view of a further
excess observed by CMS in ditau final states
at a comparable mass and similar local significance.
Moreover, an excess of events with about
$2\sigma$ local significance and consistent with a
mass of 95~GeV was observed in LEP searches for
a Higgs boson decaying to pairs of bottom quarks.
We interpret the CMS diphoton excess in
combination with the ditau excess in terms
of a pseudoscalar resonance in the
CP-conserving two-Higgs-doublet model~(2HDM).
Furthermore, we discuss the possibility
that, if CP-violation is taken into account,
a CP-mixed scalar state can
in addition describe the LEP result, thus
accommodating all three excesses simultaneously.
We find that
the region of parameter space where
both the CMS diphoton and ditau excesses can be fitted
is in tension with current constraints from
the flavour sector, potentially calling for
other new-physics contributions to flavour-physics
observables, most notably $b \to s \gamma$
transitions. We also comment on
the compatibility with the
recent ATLAS di-photon searches.
\end{abstract}
\thispagestyle{empty}
\vfill
\newpage
\setcounter{page}{1}


\tableofcontents

\section{Introduction}
\label{sec:intro}

After the discovery
of a Higgs boson at the
Large Hadron Collider~(LHC) by the ATLAS
and CMS experiments~\cite{Chatrchyan:2012xdj,Aad:2012tfa},
a prime goal of the current
LHC programme is to investigate
whether the detected Higgs boson is the only
fundamental scalar particle, according to
the predictions of the Standard Model~(SM),
or whether it is part of a
Beyond the Standard Model~(BSM) theory
with extended Higgs sectors and additional
Higgs bosons.
So far no new scalars were found at the LHC and no 
conclusive indirect hints of new physics
have been reported. With increasing precision in the measurements
of the Higgs couplings to fermions and gauge bosons,
the parameter space of BSM models has been
continuously reduced, but the present experimental
uncertainties leave room for a BSM interpretation
of the detected Higgs boson.

Many BSM theories incorporate
extended Higgs
sectors with additional scalar particles.
In particular, the presence of
additional Higgs bosons with
masses below 125~GeV is not 
excluded if their couplings
are suppressed compared to the couplings of
a SM Higgs boson.
It could very well be that these
additional Higgs bosons are within the reach of the
LHC, and 
with large enough couplings these scalars would have been produced in small numbers in past runs.
Thus, an intriguing question is whether
there could be hints for an additional Higgs boson
in the  currently existing searches
in the form
of yet non-significant excesses over the
background expectation.

In this paper, we will focus on 
a series of results presented by the CMS collaboration on searches for a scalar particle produced through gluon fusion at the LHC. These show excesses at the level of $3\sigma$ local
significance in the 
diphoton~\cite{CMS:2018cyk,CMS:2023yay,CMSnewtalk} and ditau~\cite{CMS:2022goy} final states, compatible with a scalar resonance with a mass around $95$~GeV.
Moreover, an additional mild excess of
$2\sigma$ local significance consistent with a mass
of~95~GeV was observed at the Large Electron-Positron~(LEP)
collider assuming the decay of a scalar resonance
to bottom-quark pairs~\cite{LEPWorkingGroupforHiggsbosonsearches:2003ing}.
The diphoton excess in particular has been the subject of many recent works
(see, e.g.~\citeres{Cao:2016uwt,Fox:2017uwr,
Haisch:2017gql,Biekotter:2017xmf,Liu:2018xsw,
Domingo:2018uim,Cline:2019okt,
Biekotter:2019kde,Cao:2019ofo}, where in
\citeres{Biekotter:2022jyr,Biekotter:2022abc,
Iguro:2022dok,Iguro:2022fel,
Biekotter:2023jld,Biekotter:2023qbe}
also the ditau excess
observed by~CMS was considered,
and \citeres{Biekotter:2023jld,Bonilla:2023wok}
are based on the most
recent CMS diphoton search including
the full Run~2 dataset).
We will interpret these excesses in 
the two-Higgs-doublet model (2HDM) in its charge-parity (CP)-conserving and explicitly CP-violating realization. 

Shortly after the update on
CMS's low-mass diphoton searches,
ATLAS has also published its result for
the full Run 2 search for 
low-mass Higgs bosons decaying into
diphotons~\cite{ATLAS-CONF-2023-035}.
In the ``model-dependent''
ATLAS analysis, which has very similar
experimental sensitivity to the
CMS analysis, the most pronounced excess
is observed for a
mass of 95.4~GeV, with $1.7\sigma$ local significance.
The excess is not as pronounced as the
ones reported by CMS, but the mass value
is in very good agreement, and
a possible combination of the ATLAS result
with the CMS results would give rise to slightly
smaller signal rates. 
We therefore argue that even the small upwards fluctuation seen at ATLAS cannot
exclude the signal interpretation of the~CMS diphoton excesses.
In this paper,we will solely interpret the CMS excesses but
the results can be easily extended assuming a somewhat smaller ATLAS+CMS combined
signal strength. We also wish, with this work, to show how easily a low-mass diphoton excess for 
masses of order $\sim$ 95 GeV can be fitted in the 2HDM -- if the current CMS excesses are disproven
with future observations but others appear in future data, the 2HDM should be considered as a possible
immediate explanation. 

The paper is divided as follows: \refse{sec:model} introduces the model and its CP realizations. The experimental searches that showed the excesses are discussed in detail in \refse{sec:sig},
where we also discuss the compatibility of the excesses with
corresponding ATLAS results.
Afterwards, the pertinent scenarios that accommodate the
excesses in the 2HDM are 
analised in \refse{sec:anal}, and the relevant experimental constraints are reviewed and their compatibility considered.
Further possibilities that might enable a distinction between
the realizations put forward here and other
model interpretations of the excesses published
in the past are also discussed. 
Finally, we summarize and conclude in
\refse{sec:conc}.

\section{The Model}
\label{sec:model}
\setcounter{equation}{0}

The 2HDM is one of the simplest extensions of the Standard Model of particle
physics. It was introduced in 1973 by T.D.~Lee~\cite{Lee:1973iz} to allow for extra sources of CP violation beyond the CKM matrix. The
model has the same gauge symmetries and fermionic and vector gauge content of the SM but with two $SU(2)$ hypercharge $Y=1$ Higgs doublets. Such a minimalist
extension of the SM allows for a very rich phenomenology, including the existence of
three neutral scalars and a pair of
charged scalars; violation of the CP symmetry,
both explicit or spontaneous;
dark matter candidates; non-trivial
contributions to flavour physics, via flavour-changing
neutral currents (FCNCs).
Moreover, the presence of a second Higgs doublet
is motivated by supersymmetric extensions of
the SM~\cite{HABER198575},
or by models addressing the strong CP-problem
of QCD~\cite{Kim:1986ax,PhysRevLett.38.1440}.
For a review of the 2HDM, see for 
instance Ref.~\cite{Branco:2011iw}.

The most general scalar potential of the 2HDM has 11 independent real parameters~\cite{Davidson:2005cw},
as opposed to the SM potential, which depends only on 2 real parameters. Furthermore, the fermion sector of
the most general 2HDM has twice the number of complex $3\times 3$ Yukawa matrices than the SM,
which substantially curtails the 2HDM predictive power. It is therefore customary to impose
discrete symmetries on the 2HDM Lagrangian. For instance, given that FCNC are strongly constrained by experimental results, one can consider the invariance over a
$\mathbb{Z}_2$ transformation on the doublets~\cite{Glashow:1976nt,Paschos:1976ay}, $\Phi_1 \to  \Phi_1$ and $\Phi_2 \to - \Phi_2$.
When extended to the Yukawa sector, this symmetry forces each set of fermions of the
same electric charge to couple to a single doublet, instead of both. As a consequence,
the model is free of tree-level FCNC mediated by scalars. There are four 
possibilities to extend the $\mathbb{Z}_2$ symmetry to 
the Yukawa sector, depending on how the up-type quarks, down-type quarks and charged leptons 
transform under this symmetry.\footnote{In fact there will be more
possibilities if one considers Dirac mass terms for
neutrinos, which we will not do in this paper.}
Per convention the mass of the up-type quarks is always considered to be 
proportional to $v_2$ (the vacuum expectation value of $\Phi_2$), which leaves the four possibilities
shown in Tab.~\ref{tab:types}.
\begin{table}
\centering
\begin{tabular}{rccc} \toprule
& $u$-type & $d$-type & charged leptons \\ \midrule
Type I & $\Phi_2$ & $\Phi_2$ & $\Phi_2$ \\
Type II & $\Phi_2$ & $\Phi_1$ & $\Phi_1$ \\
Lepton-specific & $\Phi_2$ & $\Phi_2$ & $\Phi_1$ \\
Flipped & $\Phi_2$ & $\Phi_1$ & $\Phi_2$ \\ \bottomrule
\end{tabular}
\caption{The four models of the $\mathbb{Z}_2$-symmetric 2HDM. Each group of 
fermions of the same electric charge is made to couple to a single doublet,
preventing tree-level FCNC interactions.
}
\label{tab:types}
\end{table}
In Type I, all fermions couple to (and gain mass from) the same doublet, 
$\Phi_2$, whereas in Type II down-type quarks and charged masses couple to $\Phi_1$
instead, a construction familiar from supersymmetric
extensions of the~SM. These choices for fermion interactions have substantial impact on the phenomenology
of each of the four models~\cite{Arbey:2017gmh}. In particular, often the most stringent constraint
from the flavour sector on these models stems from experimental data on the decay 
$b\rightarrow s\gamma$, NLO analytical expressions for which may be found 
in \citere{Enomoto:2015wbn}. For models of Type~II and Flipped, the constraints from
$b\rightarrow s\gamma$ transitions force
$\tan\beta \gtrsim 1$ and the charged Higgs-boson mass to be pushed in to the hundreds of~GeV. 
For models of Type I and Lepton-specific~(LS),
there is a lower bound on $\tan\beta$ depending on the value of the charged scalar mass.
Since in this work we will be interested in reproducing the phenomenology of a particle with a mass 
of about 95 GeV, models Type II and Flipped are therefore ruled out {\em a priori}:
a mass of the charged Higgs boson several hundreds
of~GeV
larger than the EW scale in combination with 
at least two neutral scalar states at and
below 125~GeV is incompatible with
constraints from electroweak precision measurements
(in the form of bounds on the oblique parameters
$S$, $T$ and $U$) and theoretical
constraints from perturbativity.

If the 2HDM $\mathbb{Z}_2$ symmetry is spontaneously broken, the scalar potential 
does not allow for a {\em decoupling limit},
{\em i.e.} the possibility of having a SM-like Higgs boson with mass approximately 125 GeV
and all the other extra scalars being as heavy as desired, in particular, to elude current LHC
experimental bounds~\cite{Gunion:2002zf}. 
 To contemplate the possibility of a decoupling
limit, then, one introduces a dimension-2 {\em soft breaking} term in the potential, $m_{12}^2$. The
scalar potential becomes
\beq
V &=& m_{11}^2 |\Phi_1|^2 + m_{22}^2 |\Phi_2|^2 - m_{12}^2 \left(\Phi_1^\dagger
\Phi_2 + {\rm h.c.} \right) + \frac{\lambda_1}{2} (\Phi_1^\dagger \Phi_1)^2 +
\frac{\lambda_2}{2} (\Phi_2^\dagger \Phi_2)^2 \nonumber \\
&& + \lambda_3
(\Phi_1^\dagger \Phi_1) (\Phi_2^\dagger \Phi_2) + \lambda_4
(\Phi_1^\dagger \Phi_2) (\Phi_2^\dagger \Phi_1) +
\frac{\lambda_5}{2} \left[(\Phi_1^\dagger \Phi_2)^2 + {\rm h.c.}\right] \; ,
\label{eq:pot}
\eeq
with all parameters real. The introduction of $m_{12}^2$ does not lead to any
extra infinities in high order calculations. Additionally, the scalar potential explicitly preserves CP, though
spontaneous CP breaking may occur.  On the other hand, if $m_{12}^2$ and $\lambda_5$ are complex with unrelated 
phases the CP symmetry is explicitly broken. In this paper, we will be studying both possibilities. We refer to the 
former as the {\em real} 2HDM (R2HDM), and to the latter as the
{\em complex} 2HDM (C2HDM)~\cite{Ginzburg:2002wt,Khater:2003wq,ElKaffas:2006gdt,ElKaffas:2007rq,
WahabElKaffas:2007xd,Osland:2008aw,Grzadkowski:2009iz,Arhrib:2010ju,Barroso:2012wz}. 
Another argument in favour of soft breaking the $\mathbb{Z}_2$ symmetry is to avoid
the possibility of domain walls forming~\cite{Zeldovich:1974uw}.

Regardless of soft breaking, the potential of eq.~\eqref{eq:pot} is subject to a series of basic constraints: (i) it must
be {\em bounded from below}, which means that, regardless of how large the fields become, it can never tend to minus
infinity; (ii) it must respect {\em unitarity}, so that a perturbation theory approach of quantum amplitudes
remains valid; (iii) and it must satisfy {\em electroweak precision constraints}, which stem, among others,
from precise measurements 
of the masses of the $W$ and $Z$ bosons. Constraints (i) and
(ii) impose limits on the quartic couplings of the potential (see~\cite{Ma:1978,Ivanov:2006yq,Ivanov:2007de}
and~\cite{Kanemura:2015ska,Ginzburg:2005dt}, respectively) and constraint (iii) limits mass splittings on
the scalars of the model~\cite{Haber:2010bw}. All of these bounds are implemented in the numerical scans
we will present, being a part of the code \texttt{ScannerS}~\cite{Muhlleitner:2020wwk}.
In addition, we confront the parameter points with
the cross section limits from searches for
additional Higgs bosons at LEP and the LHC
and with the LHC cross section
measurements of the detected Higgs boson at~125~GeV
using the code
\texttt{HiggsTools v.1}~\cite{Bahl:2022igd}
(which incorporates the codes
\texttt{HiggsBounds}~\cite{Bechtle:2008jh,
Bechtle:2011sb,
Bechtle:2013wla,
Bechtle:2020pkv}
and \texttt{HiggsSignals}~\cite{Bechtle:2013xfa,
Bechtle:2014ewa,
Bechtle:2020uwn}).
More details on the applied constraints
and the computation of the
respective theory predictionts
can be found in \refse{sec:constr}.

\subsection{The real 2HDM}
\label{sec:2hdm}

Though spontaneous CP breaking is possible with real $m_{12}^2$, in the current work we
are interested in a vacuum which preserves CP. Therefore, we will only consider minima
for which the doublets acquire neutral and real vacuum expectation values (vevs), $\langle \Phi_1 
\rangle = v_1/\sqrt{2}$ and $\langle \Phi_2 \rangle = v_2/\sqrt{2}$, such that 
$v_1^2 + v_2^2 = v^2 = 246^2$~GeV$^2$. Without loss of generality, we can take both vevs to be positive.
The scalar spectrum of this model yields a
pair of charged
scalar particles $H^\pm$ and three neutral
ones - two CP-even eigenstates, $h$ and $H$, and a CP-odd
one, the pseudoscalar $A$. $m_h$ and $m_H$ are the eigenvalues of a $2\times 2$ matrix, 
diagonalized by an angle $\alpha$, taken, without loss of generality, to vary between $-\pi/2$
and $\pi/2$. Another angle is defined by the ratio of the vevs, $\tan\beta = v_2/v_1$. 
The importance of $\alpha$ and $\beta$ stems from
the fact that the strength of
most of the couplings of the
scalars may, in this model, be expressed as functions of those two angles. For instance,
the couplings of $h$ and $H$ to the electroweak gauge bosons $V = W,Z$ are given by
\beq
g_{(h/H)VV}\,=\,C_{(h/H)}\,g^{SM}_{hVV}\;\;\; , \;\;\;
C_h\,=\,\sin (\beta - \alpha)\;\;\; , \;\;\;C_H\,=\,\cos (\beta - \alpha)
\label{eq:hsVV}
\eeq
where $g^{SM}_{hVV}$ is the coupling between the SM Higgs boson and the electroweak gauge bosons.
Notice how, due to the electroweak gauge symmetry, the coupling modifiers obey a sum rule,
$g_{hVV}^2 + g_{HVV}^2 = (g^{SM}_{hVV})^2$. The pseudoscalar, $A$, will not have such couplings to $W$s or $Z$s.

As for the Yukawa Lagrangian, it is given by 
\begin{eqnarray}
{\mathcal L}_\text{Yukawa} &=&
- \sum_{f=u,d,\ell} \frac{m_f}{v} \left(
\xi_h^f {\overline f}f h
+ \xi_H^f {\overline f}f H
- i \xi_A^f {\overline f}\gamma_5f A
\right)
\nonumber \\ & &
- \left\{\frac{\sqrt2V_{ud}}{v}\,
\overline{u} \left( m_u \xi_A^u \text{P}_L
+ m_d \xi_A^d \text{P}_R \right)
d H^+
+\frac{\sqrt2m_\ell\xi_A^\ell}{v}\,
\overline{\nu_L^{}}\ell_R^{}H^+
+\text{H.c.}\right\} \label{Eq:Yukawa}
\end{eqnarray}
where $m_x$ is the mass of fermion $x$, $u$ and $d$ generic up and down-type quarks respectively, and 
$V_{ud}$ the corresponding CKM matrix
element.
The coupling modifiers $\xi$ are functions of 
$\alpha$ and $\beta$ alone. In model Type I, which will be the almost exclusive focus of this
paper, they are given by 
\beq
\xi_h^f &=& \sin (\beta-\alpha)\,+\,\frac{1}{\tan\beta}\,\cos(\beta-\alpha)\,, \nonumber \\
\xi_H^f &=& \cos (\beta-\alpha)\,-\,\frac{1}{\tan\beta}\,\sin(\beta-\alpha)\,, \nonumber \\
\xi_A^u = - \,\xi_A^d &=& \frac{1}{\tan\beta}\,.
\label{eq:coumod}
\eeq
In the {\em alignment limit}, where $h$ has almost SM-like interactions, 
$\cos(\beta - \alpha)\simeq 0$, 
and we see that the magnitude of the
coupling modifiers for $H$ and $A$ grow with $1/\tan\beta$. Likewise, the strengths of the
couplings of the charged scalar increase for lower values of $\tan\beta$. As we will 
see in Section~\ref{sec:anal},
the use of these coupling modifiers allows one to considerably simplify the analysis 
of the phenomenology of the model. For the LS~model, which we will also briefly discuss
in the numerical analysis, the coupling modifiers of scalars to charged 
leptons are different than those of quarks, namely
\beq
\xi_h^l &=& \sin (\beta-\alpha)\,-\,\tan\beta\,\cos(\beta-\alpha)\,, \nonumber \\
\xi_H^l &=& \cos (\beta-\alpha)\,+\,\tan\beta\,\sin(\beta-\alpha)\,, \nonumber \\
\xi_A^l&=&  \tan\beta\,,
\eeq
now the coupling modifiers for $H$, $A$ and $H^\pm$ grow with $\tan\beta$. 
The alignment limit
has the same definition as in Type I.

For future reference, we will use the following set of input parameters to describe a R2HDM parameter point
\begin{equation}
    m_h, \quad m_H, \quad m_A, \quad m_{H^+}, \quad C_{H}, \quad \tan \beta, \quad 
    m_{12}^2 \  , \quad v  , 
\end{equation}
where $m_h=125$~GeV and $v = v_{\rm EW} = 246\gev$ are fixed to the known values.
 
\subsection{The complex 2HDM}
\label{sec:c2hdm}

If both $m_{12}^2$ and $\lambda_5$ are complex with unrelated phases, then the 2HDM scalar potential explicitly 
breaks the CP symmetry. It will not be possible to rotate all of the phases away through field re-definitions, even 
though the vevs can be set to be real and positive without loss of generality. Thus, like before, we have 
$\langle \Phi_1  \rangle = v_1/\sqrt{2}$ and $\langle \Phi_2 \rangle = v_2/\sqrt{2}$, with 
$v_1^2 + v_2^2 = v^2 = 246^2$~GeV$^2$. The now complex
soft $\mathbb{Z}_2$-breaking term, $m_{12}^2$, does not spoil the absence of tree-level FCNC. 

In this section we
review the basic notation of the C2HDM, referring
the reader to Ref.~\cite{Muhlleitner:2017dkd} for more details.
As in the real 2HDM, the scalar spectrum is composed of a
pair of charged scalars and three neutral ones. However, since the CP-symmetry
is explicitly broken, those three neutral states are not CP-even or CP-odd, rather they have indefinite CP
quantum numbers, since they result from the mixing of real and imaginary components of the neutral fields of the doublets. 
The three neutral mass eigenstates $h_i$~($i=1,2,3$) are found to be the eigenvalues of a $3\times 3$ real and 
symmetric mass matrix, diagonalized by an orthogonal rotation matrix $R_{ij}$, parameterized in terms of 
three angles $\alpha_1$, $\alpha_2$ and
$\alpha_3$ as~\cite{Barroso:2012wz}
\be
R =
\left(
\begin{array}{ccc}
c_1 c_2 & s_1 c_2 & s_2\\
-(c_1 s_2 s_3 + s_1 c_3) & c_1 c_3 - s_1 s_2 s_3  & c_2 s_3\\
- c_1 s_2 c_3 + s_1 s_3 & -(c_1 s_3 + s_1 s_2 c_3) & c_2 c_3
\end{array}
\right)\,,
\label{eq:rotmat}
\ee
where we define $s_i \equiv \sin \alpha_i$, $c_i \equiv \cos \alpha_i$. Without loss of 
generality, we can choose~\cite{ElKaffas:2007rq}
\be
- \pi/2 < \alpha_1 \leq \pi/2,
\hspace{5ex}
- \pi/2 < \alpha_2 \leq \pi/2,
\hspace{5ex}
0 \leq \alpha_3 \leq \pi/2.
\label{eq:alphas}
\ee
The angle $\alpha_2$ controls the mixture between different CP eigenstates. In particular, it may be shown
that if $s_2 = 0$ the 
scalar $h_1$ is a pure scalar;
but if $|s_2| = 1$ then $h_1$ would be a 
pure pseudoscalar.

The three $\alpha$ angles and $\beta$ appear frequently
in studies of C2HDM phenomenology.\footnote{$\tan\beta$ has the same definition in the real or complex 2HDM, since we have chosen a basis with real vevs for the C2HDM.}
For instance,
there is a relation between the masses $m_{h_i}$ of the neutral scalars and these angles, to wit
\be
m_{h_3}^2 = \frac{m_{h_1}^2\, R_{13} (R_{12} \tan{\beta} - R_{11})
+ m_{h_2}^2\ R_{23} (R_{22} \tan{\beta} - R_{21})}{R_{33} (R_{31} - R_{32} \tan{\beta})} \ ,
\label{eq:m3}
\ee
where $R_{ij}$ are the elements
of the rotation matrix shown in \refeq{eq:rotmat}.
In the same manner as in the R2HDM, the couplings of the neutral scalars to gauge bosons or fermions are
given by the respective SM coupling multiplied by a coupling modifier dependent on these angles. For the coupling
to electroweak gauge boson pairs, we have
\be
g_{h_i VV}\,=\,C_i\,g^{\rm SM}_{hVV}\;\;\; ,
\label{eq:cvv}
\ee
where~\cite{Fontes:2014xva}
\beq
C_i = c_\beta R_{i1} + s_\beta R_{i2} \;.
\label{eq:Cs}
\eeq
The sum rule mentioned in the R2HDM case still applies,
with now $C_1^2 + C_2^2 + C_3^2 = 1$. As for the 
Yukawa Lagrangian, it may be written, for a generic fermion $\psi$ of mass $m_f$, as
\beq
{\cal L}_Y = - \sum_{i=a,b,c} \frac{m_f}{v} \bar{\psi}_f \left[ c^e_i  + i c^o_i \gamma_5 \right] \psi_f h_i \;, 
\label{eq:yukc2}
\eeq
with both CP-even fermion interactions -- with coupling modifiers $c^e_i$ -- and CP-odd ones -- with 
coupling modifiers $c^o_i$.
For model Type I, the coupling modifiers are equal for all fermions, given by~\cite{Fontes:2014xva}
\be
c^e_i \,=\,\frac{1}{s_\beta}\,R_{i2}\;\;\;,\;\;\; c^o_i \,=\,\frac{1}{t_\beta}\,R_{i3}\,.
\ee
For the LS~model, the coupling modifiers
above apply only to quarks, and to charged leptons one would have
\be
c^e_i \,=\,\frac{1}{c_\beta}\,R_{i1}\;\;\;,\;\;\; c^o_i \,=\,t_\beta\,R_{i3}\,.
\ee
For future reference, we will use the following set of input parameters to describe a C2HDM parameter point
\begin{equation}
    \begin{aligned}
    &m_{h_1}, \quad m_{h_2}, \quad m_{H^\pm}, \quad \tan \beta, \quad \text{Re}(m_{12}^2), \quad v=v_{\rm EW},\\
    & C^2_2, \quad |c(h_2 t\bar{t})|^2 , \quad \mathrm{sign}(R_{23}), \quad R_{13}.
    \end{aligned}
    \label{eq:freeparasc2hdm}
\end{equation}
Using this set of input parameters,
the third neutral scalar mass $m_{h_3}$ is
a dependent parameter and computed
according to \refeq{eq:m3}.
Note that $m_{h_3}$ can smaller or larger than
$m_{h_1}$ and $m_{h_2}$, depending on the values
of the other parameters shown in \refeq{eq:freeparasc2hdm}.

\section{Possible indications for a
new scalar at 95 GeV}
\label{sec:sig}
 
In this section we briefly discuss the current
experimental status regarding several excesses observed at a mass of about 95 GeV
at LEP and LHC. We will summarize the general properties
a scalar state at~95~GeV should have such that
it can reproduce those experimental results.
For each excess considered below,
the signal strengths, or
$\mu$ values, are defined by the ratio of the observed 
number of events divided by the expected number of events
for a hypothetical SM Higss boson $\Phi_{95}$ with a mass
of 95 GeV, {\em i.e},
\be
\mu(\Phi_{95})_{X} \,=\,\frac{\sigma(\Phi_{95})\times 
\mathrm{BR}(\Phi_{95}\rightarrow X)}{\sigma^{\rm SM}(\Phi_{95})\times 
\mathrm{BR}^{\rm SM}(\Phi_{95}\rightarrow X)}\,,
\ee
where $\sigma(\Phi_{95})$ stands for the production cross
section of~$\Phi_{95}$, 
and $BR(\Phi_{95}\rightarrow X)$ is the branching
ratio for the decay of $\Phi_{95}$ into the final state $X$.

\subsection{The CMS diphoton excess}

The LHC searches for diphoton resonances
played a vital role for the discovery of
the Higgs boson at~125~GeV.
Consequently, diphoton searches are also
one of the most promising searches for
additional Higgs bosons below 125~GeV.
The CMS collaboration performed searches
for low-mass Higgs bosons decaying into
two photons at~8~TeV~\cite{CMS:2015ocq}
and 13~TeV~\cite{CMS:2018cyk}.
Combining the 8~TeV dataset from Run~1
and the first-year Run~2 data at 13~TeV,
corresponding to an integrated luminosity
of 19.7~fb$^{-1}$ and 35.9~fb$^{-1}$, respectively,
CMS observed an excess of $2.8~\sigma$ local
significance at a mass of 95.3~GeV~\cite{CMS:2018cyk}.
This excess can be described by a scalar resonance
produced in gluon-fusion production with subsequent
decay into diphotons with a signal strength
of $\mu(\Phi_{95})_{\gamma\gamma}^{\rm 1st-year} = 0.6 \pm 0.2$.
Since this excess was present
in both the 8~TeV and 13~TeV data,
it has sparked considerable attention
in the literature (see, e.g.~\bibs{Cao:2016uwt,
Fox:2017uwr,
Haisch:2017gql,
Biekotter:2017xmf,
Liu:2018xsw,
Domingo:2018uim,
Cline:2019okt}).

Recently, CMS reported an update of the
low-mass Higgs boson searches
in the diphoton final state~\cite{CMS:2023yay}.
In this new analysis, the full dataset collected
at 13~TeV (but not the Run~1 data collected
at 8~TeV) was taken into account, corresponding
to an integrated luminosity of 132.2~fb$^{-1}$. In addition, the
updated experimental analysis contains important
improvements on the background rejection from misidentified
$Z \to e^+ e^-$ Drell-Yan events, and new event classes,
demanding additional final state jets, have been used to
target different production modes.
Notably, CMS observed a local excess with $2.9~\sigma$
significance at 95.4~GeV. While the significance of the
excess and the preferred mass range
are practically unchanged compared to the previous
result,
the updated analysis is compatible
with a scalar resonance with a significantly smaller
signal strength of $\mu(\Phi_{95})_{\gamma\gamma} =
0.33^{+ 0.19}_{-0.12}$~\cite{CMSnewtalk}
as compared to the value
$\mu(\Phi_{95})_{\gamma\gamma}^{\rm 1st-year} = 0.6 \pm 0.2$
reported earlier.
 
The ATLAS collaboration reported first Run~2 results
of searches
for scalar resonances decaying into diphotons
covering the mass region below 125~GeV using
80~fb$^{-1}$ in
2018~\cite{ATLAS:2018xad}.
No excess was observed at masses around 95~GeV.
However, the resulting limits
were substantially
weaker than the corresponding CMS ones, even
at 95~GeV where CMS observed the excess.
A diphoton resonance consistent with the CMS
excess would therefore not have led to
a significant excess in this early Run~2 ATLAS search.
The final Run~2 results utilizing the full dataset
collected at~13~TeV have been reported recently
by ATLAS, just
three montsh after CMS reported its full Run~2
results~\cite{ATLAS-CONF-2023-035}.
Therein, the so-called ``model-dependent'' analysis
shows a substantially improved experimental
sensitivity compared to the earlier result,
which is now at the same level as the CMS analysis.
Notably, the most significant excess over the~SM
expectation observed by ATLAS is at~95.4~GeV,
thus in very good agreement with the masses
of the excesses observed by~CMS.
However, the excess is less pronounced, showing
a local significance of~$1.7\sigma$.
Due to the presence of the slight excess,
the ATLAS result cannot
exclude a possible diphoton signal consistent
with the excesses observed by~CMS.
However, a possible combination of both the ATLAS
and the CMS results would give rise to a slightly
smaller preferred range of the signal rate of
about~$\mu(\Phi_{95})_{\gamma\gamma}^{\rm ATLAS+CMS}
\lesssim 0.3$. Since ATLAS did not report a
signal strength value corresponding to the
observed excess, we will focus in this paper
on the description
of the excesses observed by~CMS based on the publicly
reported corresponding signal rate
of $\mu(\Phi_{95})_{\gamma\gamma} =
0.33^{+ 0.19}_{-0.12}$.
We note that the conclusions of our results would
not change significantly assuming a somewhat smaller diphoton
signal rate as suggested by the new ATLAS result. Indeed, a smaller 
diphoton signal rate at 95 GeV would even allow for a better agreement with 
flavour physics, as will be seen below.

The fact that the CMS diphoton excess was observed
with unchanged significance after the experimental
analysis was improved and after the full Run~2
dataset was considered further motivates the
investigation of possible interpretations of
such a signal in different BSM theories.
Even more importantly, however, is the change
in the required $\mu(\Phi_{95})_{\gamma\gamma}$-value.
Previously, it was shown that
the relatively large value
of $\mu(\Phi_{95})_{\gamma\gamma}^{\rm 1st-year} = 0.6 \pm 0.2$
as observed by CMS in the analysis from 2018
could not be described at the level of
$1\sigma$ in various extensions of the SM.
In particular, in the 2HDM
constraints from flavour physics in the 2HDM
push the mass scale of the BSM scalars
to masses with several hundreds of GeV in type~II and type~IV, such that
a state at 95~GeV is incompatible with
theoretical constraints
(see also \refse{sec:model}).
The scalar $H$ in the type~I 2HDM has been
considered in the past as an origin of
the CMS diphoton excess and the LEP
excess~\cite{Fox:2017uwr,Haisch:2017gql}. However, in order
to achieve sufficiently large signal rates
of $\mu(H_{95})_{\gamma\gamma}^{\rm 1st-year} \approx 0.6$ the
production of $H$ had to be enhanced
by additional contributions
such as vector-boson fusion production,
associated production with a vector boson,
or via exotic production modes via decays
of $H^\pm \to H W^\pm$.
We note that in this case it is not
clear whether the production of $H$
actually describes a signal
that is compatible with the excess
observed by CMS.
In this paper, motivated by the smaller observed
signal rate $\mu(\Phi_{95})_{\gamma\gamma} \approx 0.33$
in the updated CMS result,
we will revisit the possibility that
a BSM state at 95~GeV in the 2HDM is the
origin of the diphoton excess. We focus on such cases where this scalar
is produced dominantly
through gluon-fusion, which is the production channel
in which the diphoton excess is most pronounced,
without requiring additional production modes.
Moreover, we will consider the CP-odd scalar
$A$ as an origin of the excess, which is further
motivated by another excess observed in ditau
final states, see below.

\subsection{The CMS ditau excess}

Since Higgs bosons are expected to have
a coupling to fermions that increases with
the fermion mass, it is promising to search
for additional Higgs bosons using their decays
into third-generation fermions. In the low-mass
region below 125~GeV, the decay into top quarks
is kinematically forbidden and the decays into
bottom quarks are very difficult to search for at
the LHC due to the large QCD background.
Consequently, the searches for low-mass resonances
utilizing ditau final states are particularly important.

Low-mass Higgs-boson searches were performed
by CMS including the full Run~2 dataset collected
at 13~TeV~\cite{CMS:2022goy}.
Here it was assumed that a scalar resonance is
produced via gluon-fusion production and
$b \bar b$-associated production, and it subsequently
decays into two-lepton pairs, where final states
of both leptonically or hadronically decaying
tau-leptons were considered.
Notably, CMS observed an excess over the background
expectation assuming gluon-fusion production
at a mass compatible with 95~GeV, without a
corresponding excess assuming $b \bar b$-associated
production.
The excess was found to be most pronounced at a mass
hypothesis of 100~GeV, with local and global
significances of $3.1\sigma$ and $2.7\sigma$,
respectively. At a mass of 95~GeV, the local and
global significance amount to $2.6\sigma$ and
$2.3\sigma$, respectively, and the excess can be
described by a scalar resonance with a signal strength
of $\mu(\Phi_{95})_{\tau\tau} =
1.23^{+ 0.61}_{-0.49}$~\cite{CMS:2022goy}.
It should be noted here that the mass resolution in ditau
final states is substantially larger compared to
the mass resolution of diphoton searches, such
that a common origin of both the diphoton and the
ditau excesses by means of a scalar resonance at
95~GeV is viable although the ditau excess is
most pronounced at a mass of 100~GeV.

So far, no corresponding ATLAS searches for scalar
resonances decaying into two tau leptons exist that
cover the mass region around 95~GeV.

For a possible BSM interpretation of the ditau excess,
we stress that the CMS search specifically targets
the gluon-fusion production mode. We think that it
is therefore particularly motivated to investigate
whether a scalar state of the 2HDM can give rise
to a sizable signal in both the CMS
ditau and diphoton searches while being dominantly
produced via gluon fusion, without relying on additional
exotic production modes. In this case,
since in the 2HDM the CP-odd state $A$ at 95~GeV has
larger gluon-fusion production cross sections than
a CP-even state $H$ of the same mass, the investigation of
whether the two excesses can be described
in terms of the CP-odd Higgs boson $A$
is further motivated.
We also note that it has been recently
shown in \bib{Biekotter:2023jld} that a description
of the CMS ditau excess by means of a CP-even
resonance is in tension with the non-observation
of an excess in the CMS
searches for $t \bar t$-associated
production of a scalar resonance, with subsequent
decay into tau-lepton pairs, performed at 13~TeV utilizing the full
Run~2 dataset~\cite{CMS-PAS-EXO-21-018}.
For a pseudoscalar state,
the gluon-fusion production cross section
is enhanced due to the different Lorentz structure
of the coupling to top quarks, whereas the
cross section for $t \bar t$-associated production
is smaller as compared to a CP-even state with
similar strength of the coupling to top-quarks.
In light of this, the description of the ditau excess
via a CP-odd state is even more promising.

\subsection{The LEP excess strikes back}

Previous to the LHC, the LEP $e^+ e^-$ collider
was in operation at CERN with a
maximum center-of-mass energy of
209~GeV until the year~2000.
At~LEP, searches for the Higgs boson predicted
by the SM were performed assuming Higgsstrahlung
production with subsequent decay into pairs of
bottom-quarks and tau-leptons.
These searches excluded a SM Higgs boson up to
masses of~114.4~GeV~\cite{LEP:2003ing}.
However, Higgs bosons with masses below 114~GeV
are still
experimentally viable if their couplings to $Z$ bosons
are suppressed compared to that of a SM Higgs boson,
i.e.~$|C_i| < 1$ (see \eqref{eq:cvv}).

The LEP cross section limits extracted from searches
for scalar resonances produced in
Higgsstrahlung production with subsequent
decay into bottom-quark pairs show a mild excess
of $2.3\sigma$ local significance at a mass of
about 98~GeV~\cite{ALEPH:2006tnd}.
The excess is broad due to the hadronic
$b \bar b$ final state, such that this result is
also compatible
with a scalar resonance at 95~GeV, consistent with
the diphoton and ditau excesses observed by~CMS.
The excess can be described by a scalar resonance
with a signal strength of $\mu(\Phi_{95})_{bb} =
0.117 \pm 0.057$~\cite{Cao:2016uwt}, pointing towards
a suppression of the vector boson coupling by
roughly an order of magnitude compared to
a SM Higgs boson.

An interesting question is therefore whether
a scalar state at 95~GeV can describe also this
$b \bar b$ excess, while at the same time being
the origin of the excesses observed by~CMS.
Since the Higgsstrahlung production mechanism
requires a coupling of the scalar to $Z$ bosons,
this excess points to an interpretation via
a CP-even scalar, whereas an interpretation via a
CP-odd state is disfavoured due to the vanishing
coupling to vector bosons.
As will be discussed in more detail below,
the CP-even state $H$ of the 2HDM could give
rise to a signal compatible with the LEP excess
for values of $|\cos(\beta - \alpha)| \gtrsim
\sqrt{0.117} \approx 0.32$. Such departures
from the alignment limit are largely excluded
by the cross section measurements of the Higgs
boson at 125~GeV for all four Yukawa types of
the real 2HDM~\cite{Biekotter:2022ckj,ATLAS-CONF-2021-053}.
Even more so for the low-$\tan\beta$ range that
we will be focusing on here in order to be able to
describe the diphoton excess.

The inclusion of the LEP excess therefore
necessitates an extension of the R2HDM.
One economic possibility that was considered in
\bibs{Biekotter:2019kde,Biekotter:2021ovi,
Heinemeyer:2021msz,Biekotter:2022jyr,
Biekotter:2022abc,Biekotter:2023jld}
is to extend the 2HDM by an additional
real or complex
gauge singlet scalar field.\footnote{Interestingly,
singlet-extended 2HDMs can
additionally accommodate a sizable
upwards shift to the prediction of the $W$-boson
mass in the direction of the~CDF
measurement~\cite{CDF:2022hxs},
showing a large disagreement
with the SM prediction,
via large isospin splittings between neutral
and charged scalar states~\cite{Biekotter:2022abc}.
Moreover,
the presence of a complex singlet
field allows additionally
for the presence of a scalar dark matter candidate
at about 62~GeV whose annihilation could be
the origin of the galactic-center
excess~\cite{Biekotter:2021ovi}.}
A dominantly singlet-like
scalar state can then be the origin of the
diphoton and the LEP excess, where the suppression
of the couplings to gauge bosons is achieved
via a large singlet admixture of this state.

In this paper we take a different route, and, instead
of augmenting the particle content of the 2HDM,
we remove the restriction of CP-conservation in
the scalar potential. The motivation lies on the fact 
that scalar states with CP-mixing can potentially be
the origin of all three excesses observed at LEP and the LHC.

\subsection{Summary of possible signals}

\begin{table}[t]
\centering
\begin{tabular}
{|c|c|c|}
\hline \hline
 Channel & Signal rate & Local (global) significance \\
\hline
 & & \\
$gg \rightarrow \Phi_{95} \rightarrow \gamma\gamma$  & 
$\mu(\Phi_{95})_{\gamma\gamma} =
0.33^{+ 0.19}_{- 0.12}$~\cite{CMSnewtalk} & 
$2.9 (1.3) \sigma$~\cite{CMS:2023yay} \\
 & & \\
$gg \rightarrow \Phi_{95} \rightarrow \tau^+\tau^-$ & 
$\mu(\Phi_{95})_{\tau\tau} =
1.23^{+ 0.61}_{- 0.49}$~\cite{CMS:2022goy} &
$2.6 (2.3) \sigma$~\cite{CMS:2022goy} \\
 & & \\
$e^+e^- \rightarrow Z \Phi_{95} \rightarrow Z b\bar{b}$ & 
$\mu(\Phi_{95})_{bb} = 0.117^{+ 0.057}_{- 0.057}$~\cite{Cao:2016uwt} & 
$2.3 (<1) \sigma$~\cite{LEPWorkingGroupforHiggsbosonsearches:2003ing} \\
  & & \\
\hline \hline
\end{tabular}
\caption{Observed signal rates for a possible new 95 GeV scalar particle. 
}
\label{tab:mus95}
\end{table}

In Tab.~\ref{tab:mus95}, we summarise the current
experimental results regarding the observed
excesses in scalar particle searches with a mass close to 95 GeV. 
In this context ``scalar" stands for a spin-0 particle,
not its CP quantum numbers -- according to the
previous discussion we will indeed consider all
possible CP properties of this
particle (CP-even, CP-odd and CP-mixed).

In our numerical analysis the goal is to predict a
Higgs boson at 95~GeV whose signal rates regarding the
signatures in which the excesses were observed are in
agreement with the measured values shown within
one standard deviation, i.e.~at the level of
$1\sigma$ ore less (see Tab.~\ref{tab:mus95}). At the same time, we ensure
that the presence of the state at 95~GeV is
not in tension with
the other existing cross-section limits from searches
in which no deviations have been observed.
Note that we will not combine
the three channels in a single $\chi^2$-distribution
but instead demand in a more restrictive manner
that each excess individually is
accommodated within the uncertainty band of $\pm~1$
standard deviation.

\subsection{Additional experimental findings\
regarding a possible state at 95~GeV}

We now briefly discuss 
additional experimental results
that can be interpreted as possible hints
for a new Higgs boson at 95~GeV but which
we do not consider in our analysis since their
experimental status is less clear.

For instance,
indication for the existence of a Higgs
boson at 95~GeV was found in searches for
heavy scalar resonances decaying into the
discovered Higgs boson at 125~GeV and an
another lighter BSM Higgs boson.
CMS searched for this cascade decay assuming
that the Higgs boson at 125~GeV decays into
diphotons and the BSM Higgs boson decays into
bottom-quark pairs, giving rise to
$\gamma\gamma b \bar b$ final states,
using the full Run~2 dataset~\cite{CMS-PAS-HIG-21-011}.
An excess with local and global
$3.8\sigma$ and $2.8\sigma$ significance was found
assuming masses of 650~GeV for the heavy BSM
Higgs boson and 90~GeV for the light BSM
Higgs boson. However, CMS performed another
search with comparable sensitivity for
this cascade signature assuming that
the Higgs boson at 125~GeV
decays into a pair of tau-leptons,
giving rise to a $\tau^+ \tau^- b \bar b$
final states~\cite{CMS:2021yci}.
Therein, no corresponding
excess of events over the
SM expectation was observed
at heavy scalar masses of 600~GeV
and 700~GeV (the mass 650~GeV has not been
considered).
Given the fact that in both analysis the lighter
BSM scalar was assumed to decay into bottom-quark
pairs, and that due to
the broader mass
resolution of the $\tau^+ \tau^- b \bar b$
final state one would expect a slight excess
at masses values of 600~GeV and 700~GeV
assuming a real signal at 650~GeV consistent with
the excess in $\gamma\gamma b \bar b$,
a model interpretation of the excess in
the $\gamma \gamma b \bar b$ final state without
a corresponding signal in the
$\tau^+ \tau^- b \bar b$
final states appears to be very challenging
(see also \citere{Banik:2023ecr}).
In the 2HDM, a possible interpretation of the
excess is in addition disputed by theoretical constraints
due to the large mass splitting between 95~GeV
and 650~GeV, which can only be accommodated with
very large quartic scalar couplings.
Consequently, we will not take into account
the excess observed in \bib{CMS-PAS-HIG-21-011}
in our analysis below.

Regarding BSM cascade decays of a
heavy resonance decaying into the detected Higgs
boson at~125~GeV and another BSM scalar,
it is interesting to note that another
excess consistent with a scalar resonance
at~95GeV was observed by ATLAS in a search
for such signature in boosted hadronic
final states using novel anomaly detection
techniques~\cite{ATLAS:2023azi}.
This search is based on a fully unsupervised
machine learning analysis in order to select
possible signal events from the background-only
hypothesis, utilizing the full 13~TeV dataset
collected during Run~2.
The most significant excess for
the signature
$p p \to Y \to X h_{125} \to q \bar q b \bar b$,
with $Y$ and $X$ being the two BSM scalars,
was found for mass intervals
of $3608\gev \leq m_Y \leq 3805\gev$ and
$75.5\gev \leq m_X \leq 95.5\gev$, where
the local and global significances
of the excess amount
to about~$4\sigma$ and~$1.43\sigma$,
respectively, but the excess is,
however, not
compatible with the expected signal shape.
Evidently, as a consequence of the large mass
gap between both BSM states this excess cannot
be accommodated by means of two states of the same
SU(2) scalar doublet without violating
perturbativity constraints.
Therefore, although the presence of this additional
excess hinting at a~BSM state at~95 GeV
is certainly intriguing, 
the description of this excess requires
extending the~2HDM by additional components,
a possibility we do not consider here.

Further claims for an indication of a new
resonance at 95~GeV decaying into $W$-boson pairs
have been made in Ref.~\cite{Coloretti:2023wng}
based on a recasting of CMS~\cite{CMS:2022uhn}
and ATLAS~\cite{ATLAS:2022ooq} searches
for $h_{125} \to W^* W$. However, since the
recasting results in indications for a
new scalar with a significance above
$2\sigma$ in a wide mass range
from 90~GeV to 180~GeV, the analysis does
not point directly to the possible
presence of a state at 95~GeV.
In addition, in the 2HDM the BSM Higgs bosons
have suppressed couplings to gauge bosons,
and the decay into $W^* W$-pairs is highly
off-shell for a particle at 95~GeV. An accommodation of a signal in this final state therefore
would require very large production cross sections
which are, however, excluded based on searches
for BSM scalars in other final states.
Taking this into account, we do not consider the
possibility of an additional signal in $W^*W$ final states
in our analysis.

\section{2HDM scenarios to reproduce the 95 GeV data}
\label{sec:anal}

Having reviewed the basic definitions of the model and the possible evidence for signals
 of a new scalar with mass close to 95 GeV, we will now consider several 2HDM scenarios
to fit such signals.

\subsection{Parameter scans, theoretical and experimental constraints}
\label{sec:constr}

In order to investigate whether the 2HDM can describe the observed excesses
at 95~GeV, we generated parameter points using the code \texttt{ScannerS}~\cite{Muhlleitner:2020wwk}.
We demand the parameter points to fulfil the
following set of theoretical constraints:
the potential should be bounded from below
(BFB)~\cite{Branco:2011iw},
respect tree-level perturbative unitarity~\cite{Branco:2011iw}
and the electroweak (EW) vacuum should be the global
minimum~\cite{Barroso:2013awa,
Ivanov:2015nea}.\footnote{Since a vacuum stability analysis
is not the main goal of this paper,
we do not consider the physically
viable possibility of a long-lived metastable EW
vacuum~\cite{Biekotter:2022kgf},
and apply for simplicity the more restrictive
condition of an absolutely
stable EW vacuum.}

Additionally, we impose the following experimental constraints: we demand compatibility of the EW precision observables
in terms of the oblique parameters
$S$, $T$ and $U$ within a confidence level of
$2\sigma$~\cite{Baak:2014ora},
accounting for the experimental
correlations between the parameters,
and where the theoretical predictions
for the oblique corrections where computed
at the one-loop level according to
\citere{Grimus:2007if}.
We fix one of the scalar particles to have the mass of
the detected Higgs boson~\cite{ATLAS:2015yey},
$m_{h_{125}}=125.09~\textrm{GeV}$.
We force the mass of other scalars to be
outside of a $\pm 5$~GeV mass window around
the above value to avoid degenerate states,
in which case interference effects
would have to be taken into account.
Furthermore, we modified the code \texttt{ScannerS} to interface with the new package
\texttt{HiggsTools v.1}~\cite{Bahl:2022igd}.
Therein, we use the sub-package
\texttt{HiggsSignals}~\cite{Bechtle:2013xfa,
Bechtle:2014ewa,
Bechtle:2020uwn}
to check compatibility of the scalar
at 125~GeV with
the LHC cross section measurements.
\texttt{HiggsSignals} performs a
global $\chi^2$-fit to a large set of
LHC measurements, and based on the
$\chi^2$-value we exclude a
2HDM parameter point if it is disfavoured
compared to the SM fit result at more
than $2\sigma$ confidence level
(assuming two-dimensional parameter
estimations).
For the BSM states, the subpackage
\texttt{HiggsBounds}~\cite{Bechtle:2008jh,
Bechtle:2011sb,
Bechtle:2013wla,
Bechtle:2020pkv} 
checks the compatibility of their cross sections 
with
the~95\% confidenve-level
exclusion limits from 
searches for additional Higgs bosons.
\texttt{HiggsBounds} selects for each
BSM scalar the most sensitive search channel
based on the expected cross section limit.
For the selected channels the theoretically
predicted cross sections are compared against
the observed limits, and a point is rejected
if for one of the BSM scalars the theory
prediction is larger than the experimentally
observed limit.
Flavour-physics constraints are 
not yet imposed, but we will include a dedicated
discussion of the bounds from flavour physics below,
where we consider the compatibility with
$B\rightarrow X_s \gamma $
transitions~\cite{Deschamps:2009rh,
Mahmoudi:2009zx,Hermann:2012fc,
Misiak:2015xwa,Misiak:2017bgg, Misiak:2020vlo} in the
$m_{H^{\pm}}-\tan\beta$ plane.
In the C2HDM additional constraints from the
non-observation of electric dipole moments (EDM)
of the electron will be considered separately.
Finally, given the focus on this paper, we generate
parameter points for which
one of the BSM scalars has a mass within the mass window
$94\gev \leq m_\phi \leq 97\gev$, where in the R2DHM
we have either $\phi = A$ or $\phi = H$, and in
the C2HDM $\phi = h_1$ or $\phi = h_2$ depending on
whether $\phi$ is the lightest or the second-lightest
neutral scalar, respectively.
For both the R2HDM and the C2HDM scenarios,
we will give the scanned ranges of the
remaining input parameters in the following.

\subsection{Real 2HDM type I -- $m_A = 95$~GeV}
\label{sec:r2hdm}

In the first scenario we will assume that the
possible diphoton signal observed by~CMS is
produced by the pseudoscalar $A$, with a mass close to 95 GeV. 
We fix $m_h = 125.09$ GeV,
leaving the mass of the other CP-even scalar $H$ and of the charged Higgs to be scanned over.
To comply with bounds from searches for
charged Higgs bosons at~LEP
we set $m_{H^\pm} \geq 80$ GeV and consider additionally $m_H \geq 80$ GeV.
The scanned ranges of the complete set of
input parameters are
\begin{equation}
\begin{aligned}
m_{H}&\in[80,400] \text{ GeV}, \quad m_A \in [94,97] \text{ GeV}, \quad m_{H^\pm}\in [80,400] \text{ GeV},\\
C_H &\in [-0.3,0.3], \quad \tan \beta \in [1,3], \quad m_{12}^2 \in [10^{-3},  10^5] \text{ GeV}^2
\end{aligned}
\end{equation}
As discussed,
EW precision data constrains
the mass splittings of the scalars, which means that it will not be possible to attempt
a fit of the diphoton signal within the framework of models Type II or Flipped. In fact, in
those models flavour constraints (notably $b\rightarrow s \gamma$ bounds) force the charged 
mass to be several hundreds of GeV, and hence accommodating $m_A = 95$ GeV is impossible
when one combines both $S$, $T$ and $U$ and unitarity constraints. Therefore, we are
left to attempt a fit in models Type I or LS.
Let us leave the LS model
for later and consider Type I for now.

Identifying the 95 GeV diphoton ``signal" with a pseudoscalar means we will not be able to accommodate
the LEP $b\bar{b}$ excess -- the LEP result involved production of a scalar particle $\Phi$ via
Higgsstrahlung from a $Z$ boson, which is not possible if $\Phi$ is a pseudoscalar. Nevertheless, there are
advantages in considering a pseudoscalar when attempting a fit of the LHC diphoton result at 95 GeV: 
the main production channel is expected to be gluon-gluon fusion and that process yields,
for the same mass and similar coupling strengths,
larger cross sections for a pseudoscalar than for a CP-even scalar. 
This is due to the $\gamma_5$ matrix in
the interactions between quarks and the pseudoscalar, which enhances the quark triangle contribution
to the gluon fusion process.
Another factor enhancing the cross section in a Type I model relates to the coupling modifiers for up and
down quarks. These have opposite signs (as shown in eq.~\eqref{eq:coumod}), which induces a constructive
interference between the top and bottom loops contributing to the gluon fusion cross section 
(this effect had also been observed
in~\cite{Biekotter:2023jld}, see their footnote 2). In fact, using
\texttt{HiggsTools}, 
we obtain at N3LO QCD in the heavy-top
limit a cross section of about
76~pb 
for a CP-even SM 
scalar, i.e.~$c_i^e = 1$ and
$c_i^o = 0$, with mass of 95 GeV, and 
about
177~pb 
for a pseudoscalar of the same mass
and couplings $c_i^e = 0$ and
$c_i^o = 1$~\cite{LHCHiggsCrossSectionWorkingGroup:2016ypw}.
With a bottom coupling modifier with the 
opposite signal to the top one, this number grows to 198~pb. Since all
quark couplings of $A$ are multiplied by factors of $\pm 1/\tan\beta$, the cross section for
gluon fusion production of the pseudoscalar is therefore simply
\be
\sigma(gg\rightarrow A_{95})\,=\,\frac{198}{\tan^2\beta}\;{\rm pb}\,. 
\label{eq:sigA95}
\ee
In this scenario the phenomenology of the pseudoscalar is quite simple: with a mass of 95 GeV,  
the decays $A\rightarrow Z h$ and  $A\rightarrow Z H$ -- which would involve couplings proportional to 
$\cos(\beta-\alpha)$ and $\sin(\beta-\alpha)$, respectively -- are kinematically forbidden, as is
$A\rightarrow W^\pm H^\mp$.\footnote{Even if the final state particles in these decays are off-shell,
we can expect them to be highly suppressed.} Consequently, all possible decays of $A$ involve
fermionic couplings (even the decays $A\rightarrow \gamma\gamma$ and $A\rightarrow Z\gamma$), which
are all proportional to factors of $|1/\tan\beta|$ in Type I -- and hence the branching ratios of $A$
are independent of $\tan\beta$, or indeed of any other scalar potential parameter other than $m_A$. 
We find 
that the 
branching ratios regarding the channels
in which the excesses have been observed
are~\cite{Djouadi:1997yw,Bahl:2022igd}
\beq
{\rm BR}(A_{95}\rightarrow \gamma\gamma) &\approx&
2.8 \cdot 10^{-4} \ , \nonumber \\
{\rm BR}(A_{95}\rightarrow b\bar{b}) &\approx&
0.72 \ , \nonumber \\
{\rm BR}(A_{95}\rightarrow \tau^+\tau^-) &\approx& 0.074\,. 
\label{eq:BRAs}
\eeq
For comparison, the same branching ratios for a CP-even Higgs boson of equal mass are, respectively, 
$1.4 \cdot 10^{-3}$, $0.81$ and $0.082$.
With these numbers and those mentioned above for cross sections, 
we obtain
\be
\mu(A_{95})_{\gamma\gamma}\,=\,\frac{\sigma(gg\rightarrow A_{95})\,{\rm BR}(A_{95}\rightarrow \gamma\gamma)}
{\sigma(gg\rightarrow H^{\rm SM}_{95})\,{\rm BR}( H^{\rm SM}_{95}\rightarrow \gamma\gamma)}
\,\approx\,\frac{0.52}{\tan^2\beta}\,.
\label{eq:muAphph}
\ee
Therefore, given that the CMS results (summarised in Tab.~\ref{tab:mus95}) indicate that this quantity should 
vary between 0.21 and 0.52, we obtain the approximate allowed range for $\tan\beta$,
\be
1.00 \,\lesssim \,\tan\beta \,\lesssim\,1.57\,.
\label{eq:tb1}
\ee
For the ditau channel, we will then have
\be
\mu(A_{95})_{\tau\tau}\,=\,\frac{\sigma(gg\rightarrow A_{95})\,{\rm BR}(A_{95}\rightarrow \tau\tau)}
{\sigma(gg\rightarrow H^{SM}_{95})\,{\rm BR}( H^{SM}_{95}\rightarrow \tau\tau)}
\,\approx\,\frac{2.35}{\tan^2\beta}\,,
\ee
which, according to table~\ref{tab:mus95}, gives
a preferred $\tan\beta$ range of
\be
1.13 \,\lesssim \,\tan\beta \,\lesssim\,1.78\,.
\label{eq:tb2}
\ee
Thus, we see that the range of $\tan\beta$ necessary to fit the diphoton signal at 95 GeV intersects 
with the one required to fit the ditau signal.

\begin{figure}
  \centering
  \includegraphics[height=8cm,angle=0]{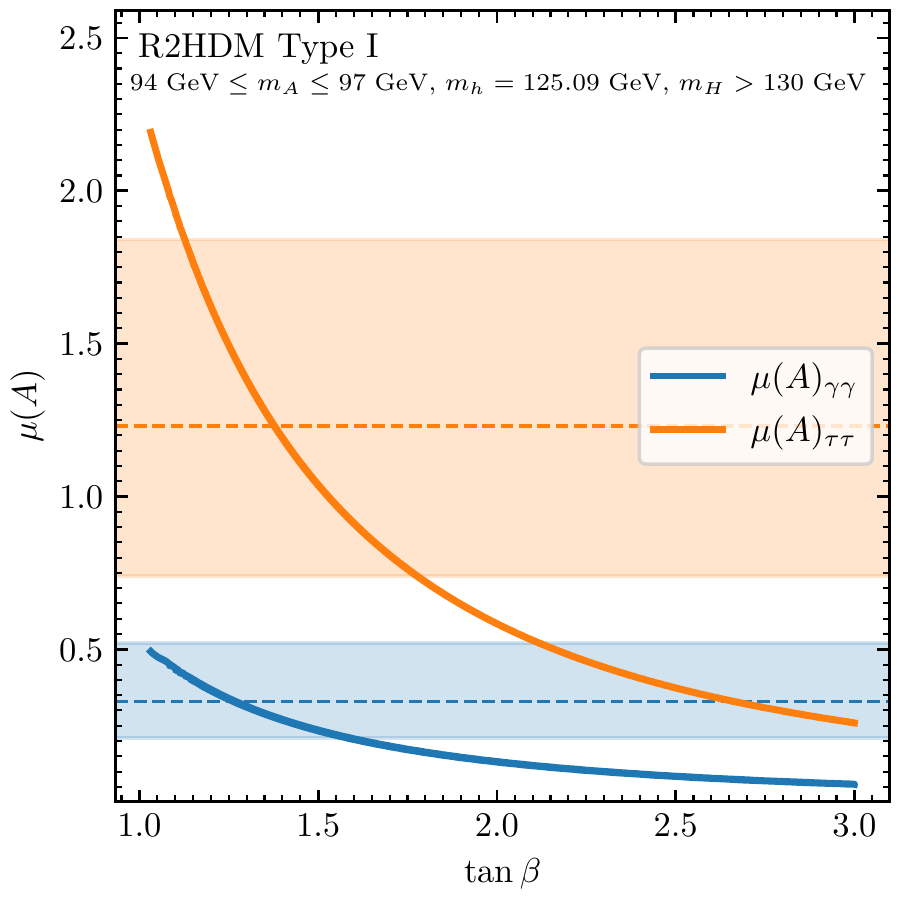}
  \caption{Signal rates for the diphoton and
  ditau excesses {\em vs} $\tan\beta$ for a
  95~GeV pseudoscalar within 
  the Type I 2HDM. Except for constraints from
  flavour physics, all
  theoretical and experimental constraints are applied.
  The bands correspond to the experimentally
  observed signal rates within their $1\sigma$
  uncertainty band (see Tab.~\ref{tab:mus95}).
  }
  \label{fig:mus_tbeta}
\end{figure}

This is shown in \reffi{fig:mus_tbeta}, where we present
the results of a scan over the real 2HDM parameter space.
Notice that 
flavour bounds were not (yet) imposed
but all others constraints (electroweak precision constraints; unitarity; boundedness from below; Higgs precision data and new scalar
search bounds) have been implemented via \texttt{ScannerS} and \texttt{HiggsTools}. From all those cuts,
we find that $m_H \in [85,370]$~GeV, $m_{H^\pm} \in [150,370]$~GeV, and will be discussed in detail
in \refse{sec:others}.
The crucial observation to be made from
\reffi{fig:mus_tbeta} is 
that both signal rates can be fitted in the range of $\tan\beta$ discussed above. 
Now, if we consider a possible combination with ATLAS, then the preferred region of the
diphoton signal rate would be constrained to  somewhat smaller values.
This would 
reduce slightly the upper end of the blue bar,
but as seen in \reffi{fig:mus_tbeta} this
would not alter the general statement that
both rates can be fitted simultaneously.

We can further characterize the region where both $\mu(A_{95})_{\gamma\gamma}$
and $\mu(A_{95})_{\tau\tau}$ are fitted. The requirement that the scalar $h$ be SM-like 
pushes us to the alignment limit, with $|\cos(\beta-\alpha)|^2 \lesssim 0.1$, see discussion
in \refse{sec:h125r}. Another interesting
observation concerns the soft breaking parameter $m^2_{12}$ -- with the low masses found for all
extra scalars, this parameter is not necessary in these scenarios. Recall that $m^2_{12}$ is introduced to 
allow for a decoupling limit, where the non-SM-like scalars can be indefinitely large. That is not the situation in this case, where a scalar at 95 GeV is set to fit the
diphoton and/or diphoton excesses. Indeed, a dedicated scan with $m^2_{12} = 0$ confirmed 
 that the conclusions reached here hold true. The possibility of fitting both excesses in the framework
 of a Type-I 2HDM with an unbroken $\mathbb{Z}_2$ is quite interesting, since in that model the
 number of free parameters in the scalar sector (7) is exactly equal to the number of 
 parameters which can be determined directly
 from experimental measurements.\footnote{These
 parameters would be the EW vev $v$ (determined from
 the gauge boson masses and the Fermi constant), 
 $\tan\beta$ (determined, e.g.~from the signal
 rates of $A_{95}$),
 $\sin(\beta-\alpha)$ (determined from the coupling
 measurements of $h$) and the four masses, $m_h$, $m_H$, $m_A$ and $m_{H^\pm}$. If
 $m^2_{12}\neq 0$,
 the determination of the model parameters
 would be more involved.}

\subsubsection{Properties of $h = h_{125}$}
\label{sec:h125r}

\begin{figure}
\centering
\includegraphics[width=0.48\textwidth]{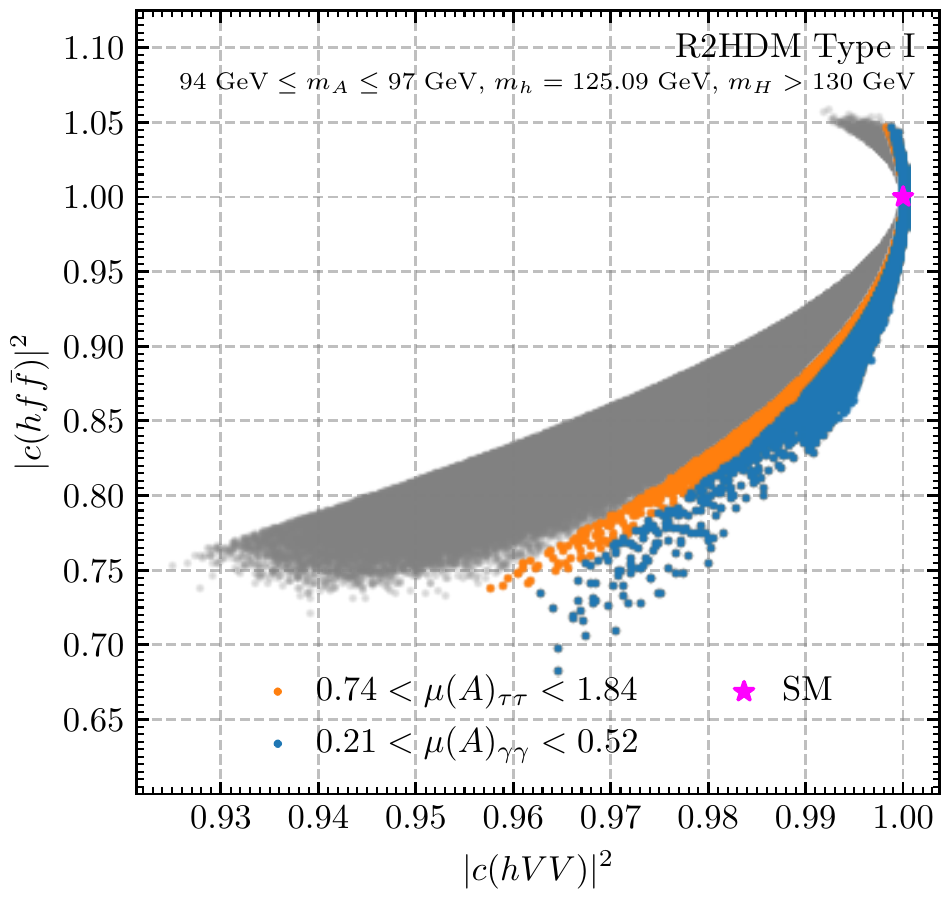}
\caption{Squared coupling coefficient of~$h = h_{125}$ to gauge bosons
$|c(h_{125}VV)|^2$
vs. to fermions $|c(h_{125}f \bar f)|^2$.
The blue and orange points predict a signal
  rate for the
  diphoton and the ditau excess within
  the experimentally observed $1\sigma$
  uncertainty bands, respectively (see Tab.~\ref{tab:mus95}).
Blue points are plotted on top of orange points.
  The remaining points are shown in grey.
The SM prediction for the couplings coefficients
is indicated with a magenta star.}
\label{fig:h125r}
\end{figure}

As already discussed in \refse{sec:intro} and \refse{sec:sig}, a wide class of
models had been proposed in the literature already
for a description of the diphoton excess at~95~GeV.
In many cases~\cite{Cao:2016uwt,Biekotter:2017xmf,Domingo:2018uim},
these models incorporate a gauge-singlet scalar
state at~95~GeV which aquires its couplings to the
fermions and gauge bosons by means of a mixing with
the detected Higgs boson at~125~GeV.
These constructions,
therefore, predict modifications of the couplings
of the detected Higgs boson compared to the SM
predictions. In the R2HDM interpretation
presented here, with the CP-odd state
being the origin of the diphoton and the ditau excess,
no mixing between $A_{95}$ and the detected Higgs
boson $h$ is required for a description of
the excesses.
Consequently, the R2HDM interpretation of the diphoton excess
and the ditau excess is possible
without the presence of a modification of the
couplings of the detected Higgs boson at~125~GeV
with respect to the SM.

This is seen in \reffi{fig:h125r}, in which we show
the squared coupling coefficients of $h$, also denoted
$\kappa$-factors in the literature,
to gauge bosons on the horizontal axis and for the to fermions on the vertical axis, respectively.
The SM prediction $\kappa^2_V = \kappa^2_f = 1$ is indicated
with the magenta star. One can see that we find points describing
the diphoton excess (blue points) and the ditau excess (orange points)
in the whole allowed range of $\kappa_f^2$. In particular, blue
and orange points are found in and around the alignment limit,
in which the state $h$ is effectively indistinguishable from
a SM Higgs boson. The allowed intervals of the $\kappa$-factors
in \reffi{fig:h125r} are constrained by the LHC cross section
measurements of the detected Higgs boson.
One can see that there is more freedom from departures of the
alignment limit for $\kappa_f < 1$ compared to the case
with $\kappa_f > 1$. The origin of this asymmetry are several LHC measurements
showing slight deviations from the SM, giving rise to a
preference for a small preference for $\kappa_f^2 < \kappa_V^2$,
(see \citere{Biekotter:2022ckj} for a more detailed discussion).
This observation is independent of the specific R2HDM scenario
under investigation here and, accordingly, not correlated with the
predictions regarding the signal rates of the pseudoscalar,
$A_{95}$.

\subsubsection{Properties of the other BSM states}
\label{sec:others}

\begin{figure}
  \centering
  \includegraphics[width=0.48\textwidth]{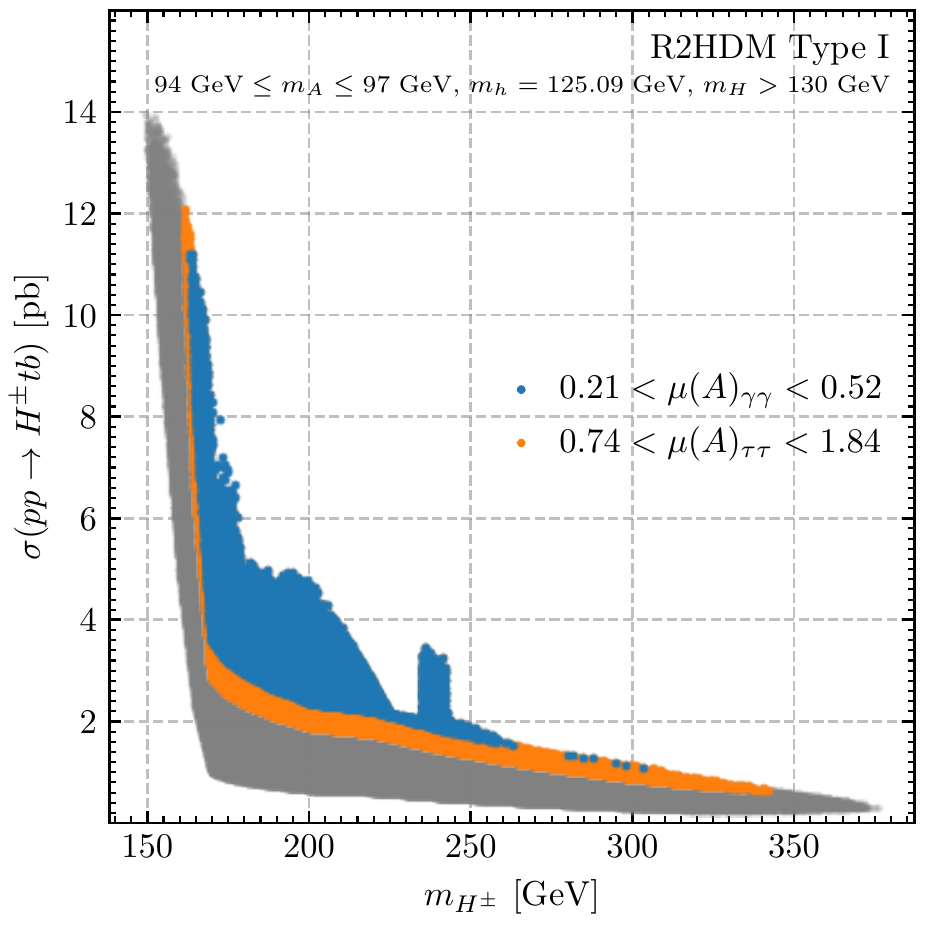}~
  \includegraphics[width=0.48\textwidth]{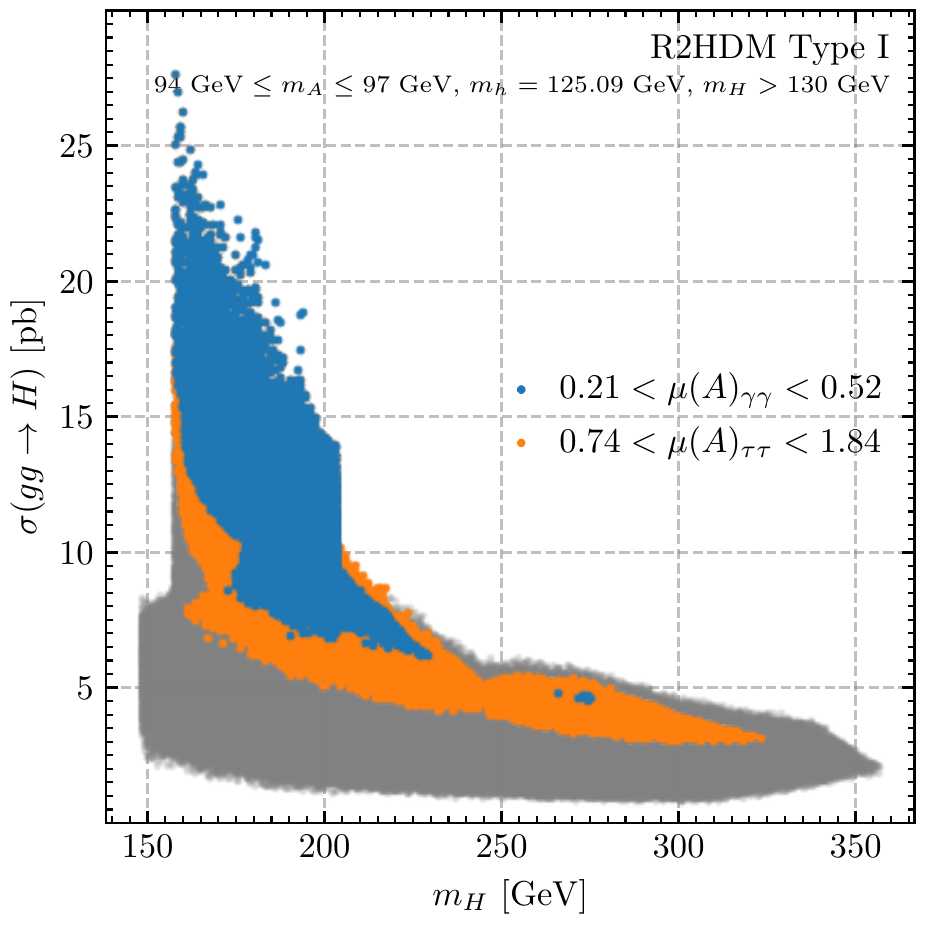}
  \caption{The cross sections $\sigma(pp \to H^\pm t b)$
  {\em vs} $m_{H^\pm}$ (left) and
  $\sigma(gg \to H)$
  {\em vs} $m_{H}$ (right)
  for the scanned points
  within the Type I 2HDM.
  Except for flavour physics, all
  theoretical and experimental constraints are applied.
  The blue and orange points predict a signal
  rate for the
  diphoton and the ditau excess within
  the experimentally observed $1\sigma$
  uncertainty bands, respectively (see Tab.~\ref{tab:mus95}).
Blue points are plotted on top of orange points.
  The remaining points are shown in grey.}
  \label{fig:mHp_xHp}
\end{figure}

More interesting are the results shown in Fig.~\ref{fig:mHp_xHp}, where we show the predicted 
values for the dominant LHC production cross sections
of $H^\pm$ (left) and $H$ (right) against their masses.
Due to the low values of $\tan\beta$
considered here, $H^\pm$ is mainly produced
in association with a top and a bottom quark,
and $H$ is mainly produced via gluon-fusion
production.
We show the points that predict values
for $\mu(A_{95})_{\gamma\gamma}$ and
$\mu(A_{95})_{\tau\tau}$
in agreement with the observed values in
blue and orange, respectively, while the
remaining points are shown in grey.
As before, all theoretical and experimental
constraints are applied, barring flavour.
We observe a clear predilection for masses
of both $H^\pm$ and $H$
lower than roughly 250 GeV, where the diphoton
excess is accommodated with cross sections
between 2~pb and 14~pb for the charged scalar,
and 5~pb and 25~pb for the heavier CP-even state $H$. 
The ditau excess by itself allows a wider range of
masses and cross sections, reflecting the fact
that a description of this excess is compatible
with larger values of $\tan\beta$ compared to the
diphoton excess (see \eqref{eq:tb2} and
\eqref{eq:tb1}).

Using \texttt{HiggsBounds}, we find that the presence of~$H^\pm$
in this scenario is mainly constrained via
searches for $pp \to H^\pm t b$ with
subsequent decay of $H^\pm \to t b$, which were
performed by the ATLAS collaboration including
the full Run~2 dataset~\cite{ATLAS:2021upq}
and by the CMS
collaboration using $35.9~\mathrm{fb}^{-1}$
collected at 13~TeV~\cite{CMS:2020imj},
all of which covering a mass range above 200~GeV.\footnote{Depending
on the experimental resolution, the limits are applied
also for masses slightly below 200~GeV.}
The blue points, which fit the CMS diphoton excess,
are limited to a relatively small region in the
mass interval of $165~\mathrm{GeV} \lesssim m_{H^\pm} \lesssim
250~\mathrm{GeV}$. This region is so far compatible with the
cross section limits from searches for $H^\pm \to tb$ but will be within the reach 
of LHC as the luminosity increases.
This provides a clear target to further probe the
R2HDM interpretation of the CMS diphoton excess
at 95~GeV in the future.
We finally note that the sharp increase of the
predicted cross sections of~$H^\pm$
at masses below about 165~GeV
results from the additional producion of $H^\pm$ via
$pp \to t \bar t$ production with the
top-quark decaying to $H^\pm$ and a bottom quark,
(see Ref.~\cite{Bahl:2021str}
for details). This increase
of the cross sections gives rise to a lower
limit of $m_{H^\pm} \gtrsim 150$~GeV due to the
limits from searches for~$H^\pm$
decaying into a tau-lepton and a
neutrino~\cite{ATLAS:2018gfm} and due to the
measurements of the branching ratios of 
ordinary top-quark decays.
In summary, only a small mass window around~200~GeV
remains allowed for~$H^\pm$, where it is too heavy to
be produced in large numbers via (off-shell) top-quark
decays, but also light enough to escape the LHC
searches for $H^\pm \to tb$.

Turning now to the second CP-even Higgs boson~$H$,
one can see in the right plot of Fig.~\ref{fig:mHp_xHp}
that parameter points describing the CMS diphoton excess
(blue points) can be found only in a narrow mass
interval of $170~\mathrm{GeV} \lesssim m_H \lesssim
225~\mathrm{GeV}$. Within this interval, the CMS searches for
$H$ decaying into a pair of
$Z$-bosons including $35.9~\mathrm{fb}^{-1}$
collected at 13~TeV~\cite{CMS:2018amk}
give rise 
to the upper limits on the gluon-fusion production
cross sections of~$H$.
For masses of $m_H \gtrsim 200\gev$, we observe a sharp
decrease of the allowed values of the cross sections,
almost excluding the possibility of a description
of the diphoton excess. This decrease of the maximal
allowed values of $\sigma(gg \to H)$ result from searches
for the cascade decay $H \to Z A$, where here $A = A_{95}$ is
the state responsible for the description of the excesses
at 95~GeV (see also \citere{Iguro:2022fel}),
which have been performed
by the CMS collaboration at~8~TeV assuming that
$A$ decays into a pair of bottom quarks or a pair
of tau leptons~\cite{CMS:2016xnc}, and at 13~TeV assuming
$A \to b \bar b$~\cite{CMS:2019ogx}. We note that the
corresponding ATLAS analysis, including already the
full Run~2 data set, does not cover the mass region
below 125~GeV for the state $A$, and is therefore
not applicable here.
The lower limit on~$m_H$, for which blue (and orange)
points are found, is determined by the CMS cross section
limits from searches for scalar resonances decaying
into tau lepton pairs, which have been performed
at 13~TeV using the full Run~2 data
set~\cite{CMS:2022goy}.
This search becomes the most sensitive
channel for the presence of~$H$ at masses below the
kinematic threshold for the decays $H \to VV$, where
the branching ratios for $H \to \tau^+ \tau^-$ become sizable.
Again, we note that the corresponding ATLAS search
for scalar resonance decaying into tau-lepton pairs
is not applicable here because light masses below 200~GeV
are not covered in the analysis~\cite{ATLAS:2020zms}.

\begin{figure}
  \centering
  \includegraphics[width=0.48\textwidth]{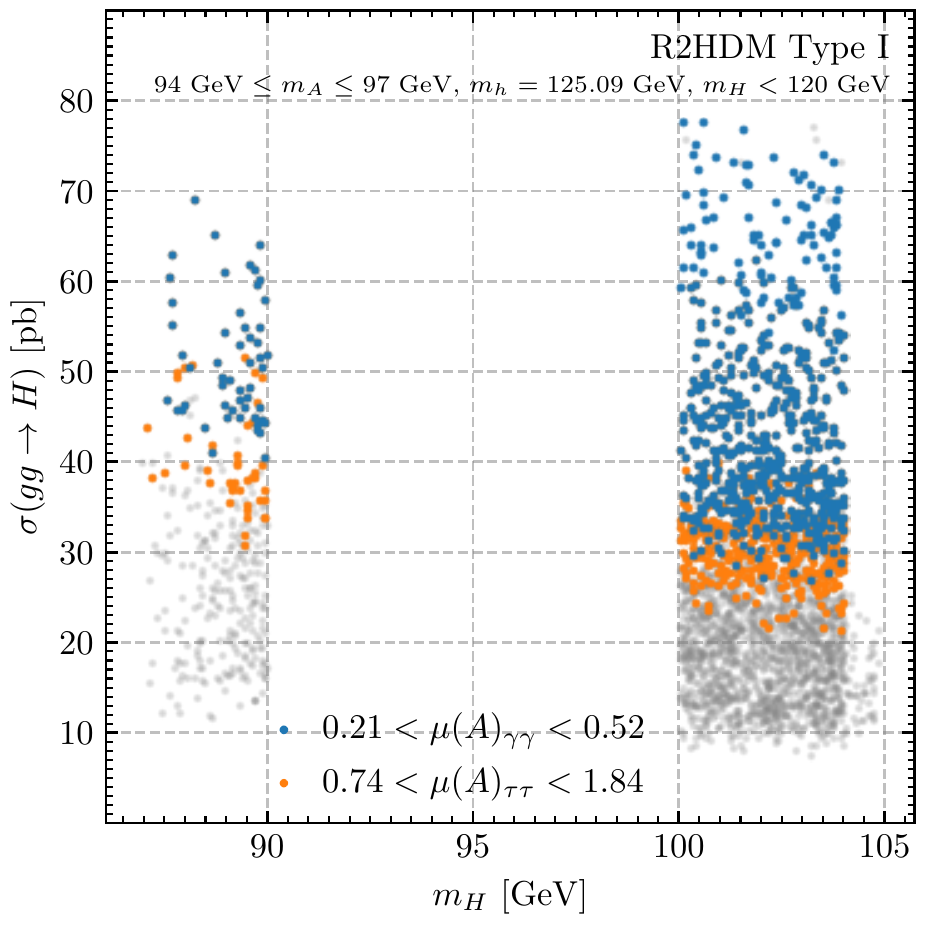}
  \caption{Same as Fig.~\ref{fig:mHp_xHp} but showing
  all scan points with $m_H < 120\gev$.
  The mass window $90\gev < m_H < 100\gev$ was excluded
  in our analysis to avoid degenerate neutral scalar states,
  see discussion in the text.}
  \label{fig:mHp_xHp_low}
\end{figure}
As noted, we also find parameter points
where there is a
second CP-even scalar with a mass below 125~GeV. We focus on the case where
we force non-degeneracy of masses between the scalars. We show the corresponding 
results in \reffi{fig:mHp_xHp_low}, where one can observe
that the cross section can grow up to values of
about 80~pb for masses of~$H$ slightly above 100~GeV.
The most sensitive search channels for most
parameter points are LHC searches for the decay
$H \to \tau^+ \tau^-$ at~13~TeV, which were
published in \citere{CMS:2018rmh,CMS:2022goy}
assuming that~$H$ is produced via gluon-fusion,
and in \citere{CMS-PAS-EXO-21-018}
assuming the production of~$H$ in association
with top-quark pairs.
The masses of the charged Higgs bosons in this
scenario are found in the range $140\gev \lesssim m_{H^\pm}
\lesssim 175\gev$, where the phenomenology is unchanged
compared to the case with $m_H > 130\gev$ as discussed in
\refse{sec:others}.

Interestingly,
the parameter points with $m_H \approx 100\gev$
open up a further possibility to describe the
excesses observed by CMS.
We find that in this
case the cross sections for the process
$gg \to H \to \tau^+ \tau^-$ can be large enough
to accommodate the ditau excess (which actually
was most pronounced at~100~GeV), while the diphoton
excess is accommodated by means of the pseudoscalar
state~$A$. In order to accurately determine whether
the ditau excess is described in agreement with
the experimental observations, one would have to
study the impact of both the contributions of~$A$
and~$H$ to the possible ditau signal, and verify whether
the observed excess is sufficiently broad that
one can sum both contributions (where interference
effects do not play a role if CP-conservation is imposed),
or whether both states would be observed as two
distinguishable signals in the ditau invariant-mass spectrum.
Since we focus here on the possibility to accommodate
the collider excesses by means of a single BSM
particle, which we regard as a theoretically more
economical description providing more specific
predictions to probe the scenarios in the future,
we do not study in detail the case with two particles contributing
to the possible ditau signal.

\subsubsection{Flavour-physics constraints}

\begin{figure}
  \centering
  \includegraphics[height=8cm,angle=0]{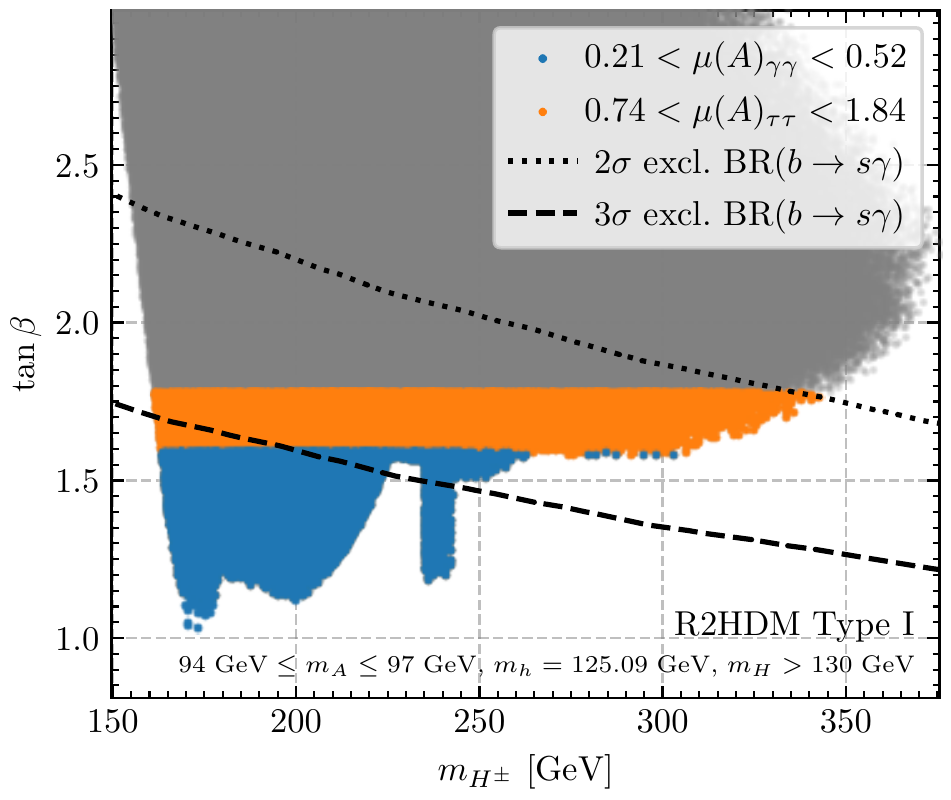}
  \caption{Flavour constraints in the $m_{H^\pm}$--$\tan\beta$ plane.
  The regions below the
  black dotted and dashed lines
  are excluded at $2\sigma$ and $3\sigma$ confidence level,
  respectively. 
  Grey points are the totality of our scan; orange points are
  restricted to  $0.74<\mu(A)_{\tau\tau}<1.84$;
  and blue points are further restricted to
  $0.21<\mu(A)_{\gamma\gamma}<0.52$. }
  \label{fig:fl_A95}
\end{figure}

Finally, let us consider the impact of flavour constraints in these results. For a Type I model, the
strongest bounds on the 2HDM parameter space -- usually expressed as excluded areas in the 
$\tan\beta$--$m_{H^\pm}$ plane -- come from $b\rightarrow s \gamma$ measurements~\cite{Arbey:2017gmh}.
Almost as restrictive are the bounds from $\Delta m_{B_s}$, the mass difference between the $B_s$ 
and $\bar{B}_s$ neutral mesons. In Fig.~\ref{fig:fl_A95}, we display the totality of our scan in the
$\tan\beta$--$m_{H^\pm}$ plane, where the regions below the
black dotted and dashed lines
are excluded at $2\sigma$ and $3\sigma$ confidence level,
respectively. These lines were obtained using
\texttt{SuperIso}~\cite{Mahmoudi:2007vz,Mahmoudi:2008tp}
for the theoretical predictions for $\mathrm{BR}(b \to s \gamma)$,
and comparing them to the experimentally measured central
value and $\pm~1$ standard deviations shown in
\refeq{eq:flavexp}.
Blue points 
predict a value of~$\mu(A_{95})_{\gamma\gamma}$
within the $1\sigma$ uncertainty interval.
In the same manner, orange points predict
a value of~$\mu(A_{95})_{\tau\tau}$ compatible
with the measured value within $1\sigma$.
We recognise that the blue points, providing
a good description of the diphoton excess, are
disfavoured by the flavour constraints at
a confidence level of about~$2.5\sigma$ or more.
We note that this tension would be weaker
to some degree
assuming a combined ATLAS+CMS diphoton signal rate,
which would be compatible with slightly larger
$\tan\beta$~values.
 
In order to further quantify this, let us recall that there are considerable uncertainties in the calculation of 
the $b\rightarrow s \gamma$ branching ratio already at the SM, stemming from choices of renormalization or 
fragmentation scales. The SM prediction for this branching ratio (see~\cite{Misiak:2020vlo}) 
gives
\be
 {\rm BR}(b \to s \gamma)_{E_\gamma > E_0=1.6~{\rm GeV}} = (3.40 \pm 0.17) \times 10^{-4}\,,
\ee
 and the experimentally measured value for this quantity from the HFLAV 
 Collaboration~\cite{HeavyFlavorAveragingGroup:2022wzx} is
\be 
 {\rm BR}(b \to s \gamma)_{E_\gamma > E_0=1.6~{\rm GeV}} = (3.49 \pm 0.19) \times 10^{-4}\,.
 \label{eq:flavexp}
\ee
In the 2HDM, the $b\to s\gamma$ branching ratio includes the SM value plus a deviation $\delta$ containing
charged Higgs contributions~\cite{Enomoto:2015wbn}, 
\be
{\rm BR}(b \to s \gamma)_{E_\gamma > E_0}= {\rm BR}(b \to s \gamma)^{\rm SM}+\delta {\rm BR}(b \to s \gamma)\,,
\ee
so that the quantity $\delta {\rm BR}(b \to s \gamma)$ must be smaller than some combination of the
theoretical and experimental errors above. If one takes
$|\delta {\rm BR}(b \to s \gamma)|<2.5\times 10^{-5}$, this reproduces with very good approximation the exclusion
region from~\cite{Arbey:2017gmh}. Using the NLO expressions
from~\cite{Enomoto:2015wbn}\footnote{NLO contributions
account for a 15\% to 20\% correction on the LO result.},
we find that the blue or orange points from \reffi{fig:fl_A95} yield
$|\delta {\rm BR}(b \to s \gamma)| \simeq (4-14)\times 10^{-5}$. Therefore, unless a non-2HDM contribution
to the flavour sector is introduced to account for this divergence, the interpretation of the diphoton 
excesses at 95~GeV as a Type-I pseudoscalar $A$ is in tension with the current experimental
$b\to s\gamma$ results.
It should be added that it is possible to satisfy the $b\to s\gamma$ bound in a larger mass regime: with $m_H$ 
and $m_{H^\pm}$ above about 500~GeV, it is possible to accommodate the values of $\tan\beta$ required to fit both
$\mu(A_{95})_{\gamma\gamma}$ and $\mu(A_{95})_{\tau\tau}$. However, in that regime there are strong constraints from
searches in the $H\to ZA \to llb\bar{b}$ channel~\cite{CMS:2019ogx} -- notice that the vertex $HZA$ 
is proportional to $\sin(\beta-\alpha) \simeq 1$ -- so the decay $H\to ZA$ is not at all suppressed despite 
proximity to the alignment limit. And with $H$ masses above 500 GeV and $m_A = 95$ GeV, phase-space suppression
does not affect this decay either.
Moreover, masses of $250~\mathrm{GeV} \lesssim m_{H^\pm}
\lesssim 350~\mathrm{GeV}$
with $\tan\beta < 2$ (as favoured by the excesses)
are excluded by LHC searches for $H^\pm \to tb$ (see Fig.~\ref{fig:mHp_xHp}) and even larger values
of $m_{H^\pm}$ are in tension with theoretical constraint
from perturbative unitarity due to the large mass splitting
between $A_{95}$ and $H^\pm$.

\subsection{Real 2HDM type I -- $m_H = 95$~GeV}

Another obvious possibility within the R2HDM would be to attempt to explain the CMS di-photon
excess as being produced by the CP-even scalar $H$. This would also raise the possibility of a simultaneous
explanation of the LEP excess, since $H$ could have been produced via Higgstrahlung at LEP. However,
a dedicated scan of this possibility -- fixing $m_H \simeq 95$ GeV and allowing $m_A, m_{H^\pm} > 130\gev$ to 
vary freely -- leads to no satisfactory conclusions. To wit, the only way of having values of $\mu(H)_{\gamma\gamma}$
above 0.21 would be to have $\tan\beta\lesssim 0.6$, and this could only be achieved in the~LS model.
In that case the CMS di-tau excess could be accommodated for $\tan\beta\geq 1$, incompatible
with the $\tan\beta$-values required for a description of the diphoton excess.
In a Type I model the results are even worse, with maximum values of $\mu(H)_{\gamma\gamma}$ of the order of 0.03
obtained for $\tan\beta \simeq 1.5$, and $\mu(H)_{\tau\tau}$ smaller than 0.6 for all scan points. In both cases, 
no possibility of reproducing the LEP excess was found
due to the constraints on the signal rates of
the Higgs boson at~125~GeV, setting tight upper limits
on the strength of the coupling of~$H$ to
$Z$~bosons due to the sum rule discussed
below \refeq{eq:hsVV}.
Taking all these limitations into account, we
do not discuss the scenarios with
$m_H \simeq 95\gev$ any further.

\subsection{Real 2HDM type LS}

\begin{figure}
  \centering
  \includegraphics[height=8cm,angle=0]{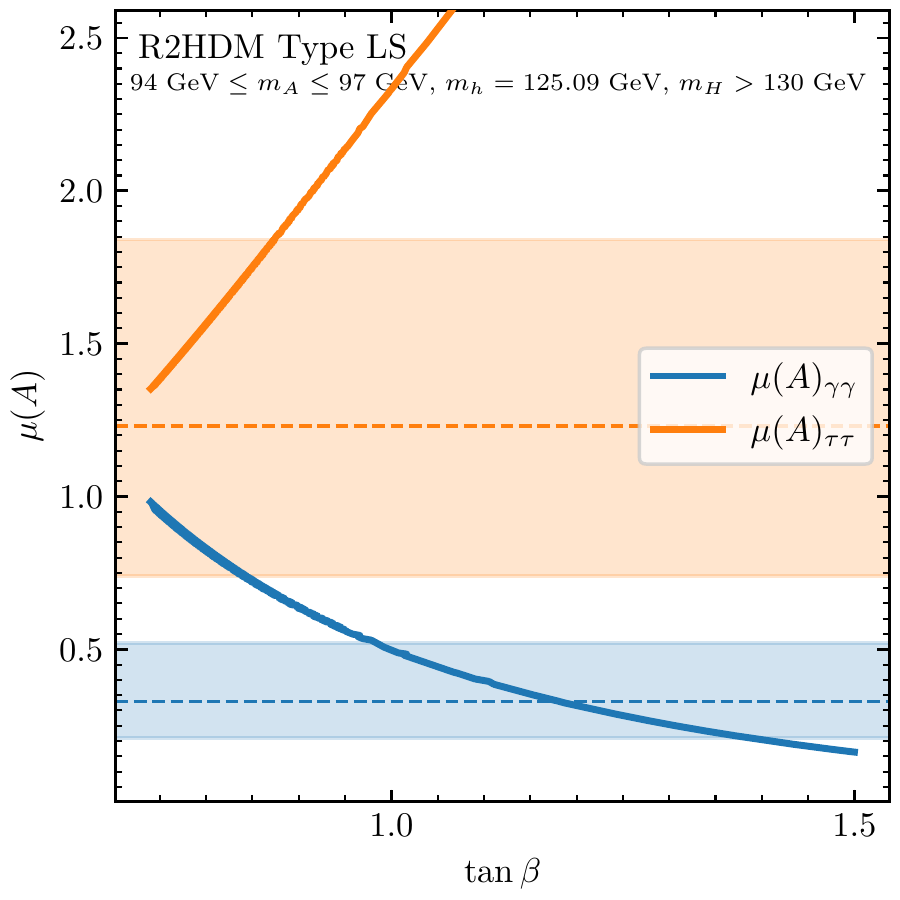}
  \caption{Signal rates for the diphoton and
  ditau excesses {\em vs} $\tan\beta$ for a
  95~GeV pseudoscalar within 
  the Lepton Specific 2HDM. Except for constraints from
  flavour physics, all
  theoretical and experimental constraints are applied.
  The bands correspond to the experimentally
  observed signal rates within their $1\sigma$
  uncertainty band (see Tab.~\ref{tab:mus95}).
  }
  \label{fig:mus_tbetaLS}
\end{figure}

Besides a Type I Yukawa model, the lepton specific (LS) 2HDM also
allows for light scalar spectra with masses around $125\gev$
without strong disagreements with flavour-physics constraints,
measurements of EWPO, or theoretical requirements of perturbativity.
Here, we briefly comment on the possibility of describing the
excesses at~95~GeV in the LS R2HDM.
We can obtain a semi-analytical understanding of the
predictions of the LS model for 
$\mu(A_{95})$. Since only the leptonic coupling modifier changes compared to Type I, the expression 
of the cross section on $\tan\beta$ shown in~\eqref{eq:sigA95} is the same for both models. From~\eqref{eq:BRAs}, 
we also see that $BR(A\to \tau\tau)$ is already small compared to other decay channels. The branching ratios
for $A$ in the LS model now have a non-trivial $\tan\beta$ dependence -- the leptonic decay
widths are proportional to $\tan^2\beta$ -- all others proportional to $1/\tan^2\beta$\footnote{Neglecting the leptonic contributions to
$A\to \gamma\gamma$, which is a good first approximation.}. With $\Gamma_A$
being the total width of the pseudoscalar $A$, we can use the numbers of \refeq{eq:BRAs} and single out 
expressions for several widths with their explicit $\tan\beta$ dependences,
\begin{align}
\Gamma(A\to \gamma\gamma) &\simeq \, \displaystyle{\frac{2.8\times 10^{-4}}{\tan^2\beta}}\,\Gamma_A\,,\nonumber \\
\Gamma(A\to \tau\tau) & =  \, 0.074\,\tan^2\beta\,\Gamma_A\,, \nonumber \\
\Gamma(A\to \rm{Not}\,\,\tau\rm{'s}) & \simeq \displaystyle{\frac{1 - 0.074}{\tan^2\beta}}\,\Gamma_A\,.
\end{align}
With these approximate expressions, we can write
\be
BR(A\to \gamma\gamma) \,= \, \frac{\Gamma(A\to \gamma\gamma)}{\Gamma(A\to \rm{Not}\,\,\tau\rm{'s}) + \Gamma(A\to \tau\tau)} \,\simeq \, \frac{2.8\times 10^{-4}}{0.926\,+\,0.074\,\tan^4\beta}
\ee
and likewise we obtain
\be
BR(A\to \tau\tau) \,= \, \frac{\Gamma(A\to \tau\tau)}{\Gamma(A\to \rm{Not}\,\,\tau\rm{'s}) + 
\Gamma(A\to \tau\tau)} \,\simeq \,\frac{0.074}{0.926\,+\,0.074\,\tan^4\beta}\,\tan^4\beta\,.
\ee
We therefore see that both branching ratios have very different $\tan\beta$ dependences, which leads us to 
(using the cross section from~\eqref{eq:sigA95})
\bea
\mu(A_{95})_{\gamma\gamma}&=&\displaystyle{\frac{0.52}{(0.926\,+\,0.074\,\tan^4\beta)\,\tan^2\beta}}\,\nonumber \\
\mu(A_{95})_{\tau\tau}&=&\displaystyle{\frac{2.35\,\tan^2\beta}{0.926\,+\,0.074\,\tan^4\beta}}
\eea
which reproduce quite accurately the results of our scan, shown in Fig.~\ref{fig:mus_tbetaLS}. As we 
see from this plot, there is no region of $\tan\beta$ values for which both diphoton and ditau
signals can be fitted simultaneously. The ditau signal can be fitted, but only for values
of $\tan\beta$ below 1, and the diphoton signal once again requires $1.0\lesssim\tan\beta\lesssim 1.3$.
In both cases, as in the Type I 2HDM, such small $\tan\beta$-values
are in significant tension with constraints from flavour physics.
Since the LS Yukawa type does not allow for a description
of the diphoton excess, which is the main motivation for
this paper, as discussed above, we will not consider
the LS~type in the following discussion.

\subsection{Complex 2HDM -- $m_{h_1} = 95$ GeV, $m_{h_2} = 125$ GeV, $m_{h_3} > 130$ GeV}
\label{sec:mhe}

As we have seen in the previous section,
it is possible to  describe well
the 95 GeV diphoton and ditau excesses
via a pseudoscalar within the
CP-conserving Type-I 2HDM but the corresponding parameter space
was found to be in tension with flavour-phyics constraints,
namely the $b\to s \gamma$ bounds. Furthermore, 
a pseudoscalar cannot serve as an
origin for the LEP $b\bar{b}$ excess.
A CP-even scalar with 
mass equal to 95~GeV cannot describe the diphoton
excess in this model without even more strenuous violations
of flavour bounds due to the 
 low values for $\tan\beta$ required.
It is, therefore, of interest
to attempt a description of the excesses
within the wider framework of the C2HDM.
In this model, all scalars have in general
mixed CP-properties
and one can envision a 95~GeV particle with a
significant CP-odd component, by means of
which the particle can account for 
the LHC diphoton and ditau excesses,
and a CP-even component large enough to
produce the possible $b\bar{b}$ signal at~LEP.

Other possible advantages arising from the C2HDM are:
(a) a widening of the available parameter space, due
to an extra parameter in the model; (b) greater flexibility in complying with Higgs precision data, since the
gluon-gluon production cross section now includes contributions from both CP-even and CP-odd
channels; (c) possible larger values for the branching ratios for the diphoton decay of the 95 GeV particle
than would be obtained for a pure pseudoscalar.\footnote{In fact, 
for a CP-odd state of mass 95~GeV
($c^e_i = 0$ and $c^o_i = 1$)
the branching ratio for the diphoton decay
is $2.8\times 10^{-4}$,
whereas for a CP-even SM-like state
($c^e_i = 1$, $c^o_i = 0$) of the same mass
one finds $1.4\times 10^{-3}$.}

We, therefore, undertook a parameter scan of the C2HDM, following the procedures outlined in Sect.~\ref{sec:model},
including all theoretical (boundedness from below; unitarity; vacuum stability) and experimental constraints (electroweak
precision bounds; Higgs precision data; extra scalar searches), with the exception of flavour
physics and electron dipole moment (EDM)
measurements. In the C2HDM, the latter may
become relevant~\cite{Jung:2013hka}, and
we will address both separately in
the following discussion. 
The scanned ranges of the complete set of
input parameters of the C2HDM are
\begin{equation}
\begin{aligned}
&m_{h_1}\in [94,97]\gev, \quad m_{h_2} = 125.09\gev, \quad m_{H^\pm} \in [80,400]\gev, \\
&\text{Re}(m_{12}^2)\in [10^{-3}, 10^5 ]\gev^2, \quad C_2^2\in[0.75,1], \quad |c(h_2 u\bar{u})|^2 \in [0.8,1.2], \quad \tan \beta \in [1,3],\\
&R_{13}\in [-1,1], \quad \text{Sign}(R_{23})\in[-1,1]
\end{aligned}
\label{eq:parasc}
\end{equation}
According to~\eqref{eq:m3}, specifying two of the neutral masses,
$\tan\beta$ and the mixing matrix $R$, the third mass
$m_{h_3}$ is fixed by the parameters shown in \refeq{eq:parasc}. There are then three possibilites, 
(i) $m_{h_3} > 125$ GeV, (ii) $95 < m_{h_3} < 125$ GeV and (iii) $m_{h_3} < 95$ GeV. Option (i) seems the 
most natural way to accommodate a scalar with 95 GeV, since options (ii) and (iii) would imply the existence
of two lighter scalars than the SM-like state with mass 125 GeV discovered at the 
LHC.\footnote{And paraphrasing Wilde, to lose one scalar may be regarded as a misfortune; to lose both 
looks like carelessness. }
We will therefore 
deal with option (i) for this section,
and briefly address the other two in Sect.~\ref{sec:mli}.

\begin{figure}[t]
\centering
\includegraphics[height=8cm,angle=0]{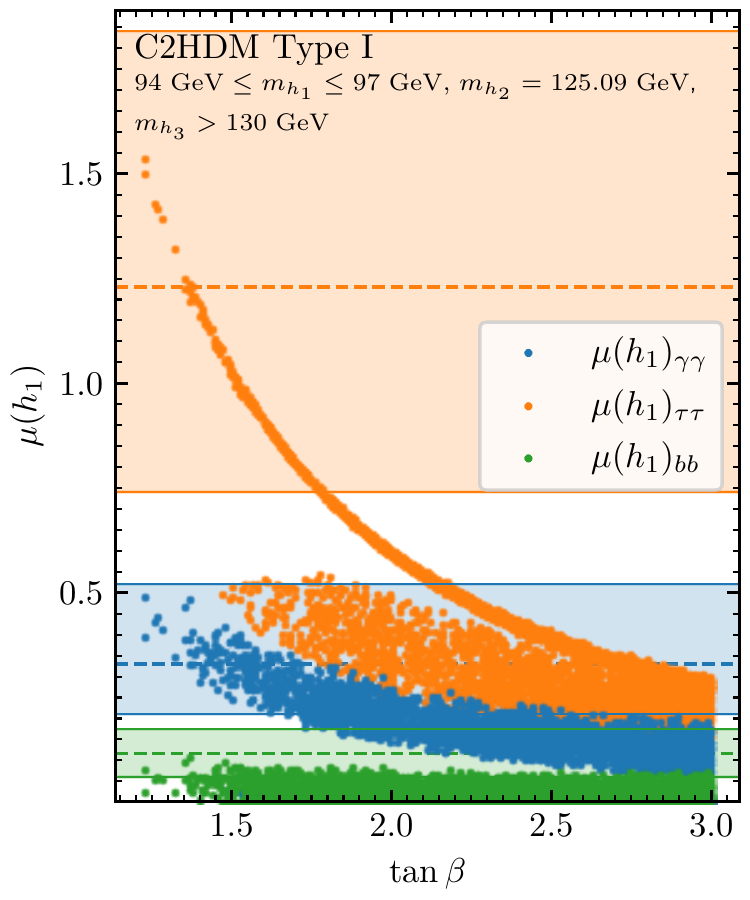}
  \caption{Signal rates for the diphoton
  (blue), ditau (orange) and $b \bar b$ (green)
  excesses {\em vs}
  $\tan\beta$ for a 95~GeV scalar within the
  Type~I C2HDM and $m_{h_3} > 130\gev$.
  Except for constraints from flavour physics and EDMs, all
  theoretical and experimental constraints are applied.
  The bands correspond to the experimentally
  observed signal rates within their $1\sigma$
  uncertainty band (see Tab.~\ref{tab:mus95}).}

\label{fig:tbeta_mus_c2hdm}
\end{figure}

The results of our C2HDM scan in the mass
hierarchy with
$m_{h_1} \approx 95\gev$,
$m_{h_2} = 125.1\gev$ and
$m_{h_3} > 130$ GeV (we only consider non-degenerate scenarios, see Sect~\ref{sec:constr})
are shown in \reffi{fig:tbeta_mus_c2hdm},
where we plot, as a function
of $\tan\beta$, the values of the signal rates $\mu(h_1)_{\gamma\gamma}$,
$\mu(h_1)_{\tau\tau}$ and $\mu(h_1)_{bb}$ (the latter concerning LEP results, the former two LHC ones)
in blue, orange and green, respectively.
We see that the diphoton excess can be reproduced for $1.0 \lesssim \tan\beta \lesssim 2.5$, a wider 
interval than the one found for the real 2HDM
(see \reffi{fig:mus_tbeta}).
A joint fit with the ditau signal reduces the upper
bound on $\tan\beta$ to roughly 1.8.
As for the LEP excess, we see that the lower values of the
$1\sigma$ interval for this signal are reproduced in the C2HDM more or less independently of
$\tan\beta$ (at least for the restricted scan we undertook here). Remarkably, we see that for
$1.0 \lesssim \tan\beta \lesssim 1.8$ all
three possible 95~GeV signals may be
described individually in the C2HDM.
Even more strikingly,
the parameter points with $\tan\beta \lesssim 1.5$
predict signal rates regarding all three excesses within
the respective experimentally observed
$1\sigma$ uncertainty bands, indicating that the
excesses can be fitted simultaneously in the C2HDM,
in contrast to the observation in the R2HDM
in which no sizable signal at LEP was present.
Again, a possible combination 
of the di-photon search results of both CMS and ATLAS
does not alter 
the general statement.
If the diphoton signal rate would be constrained to
slightly smaller values, as the ATLAS result suggests,
the three excesses
could still be fitted simultaneously.

\begin{figure}
\centering
\includegraphics[height=8cm,angle=0]{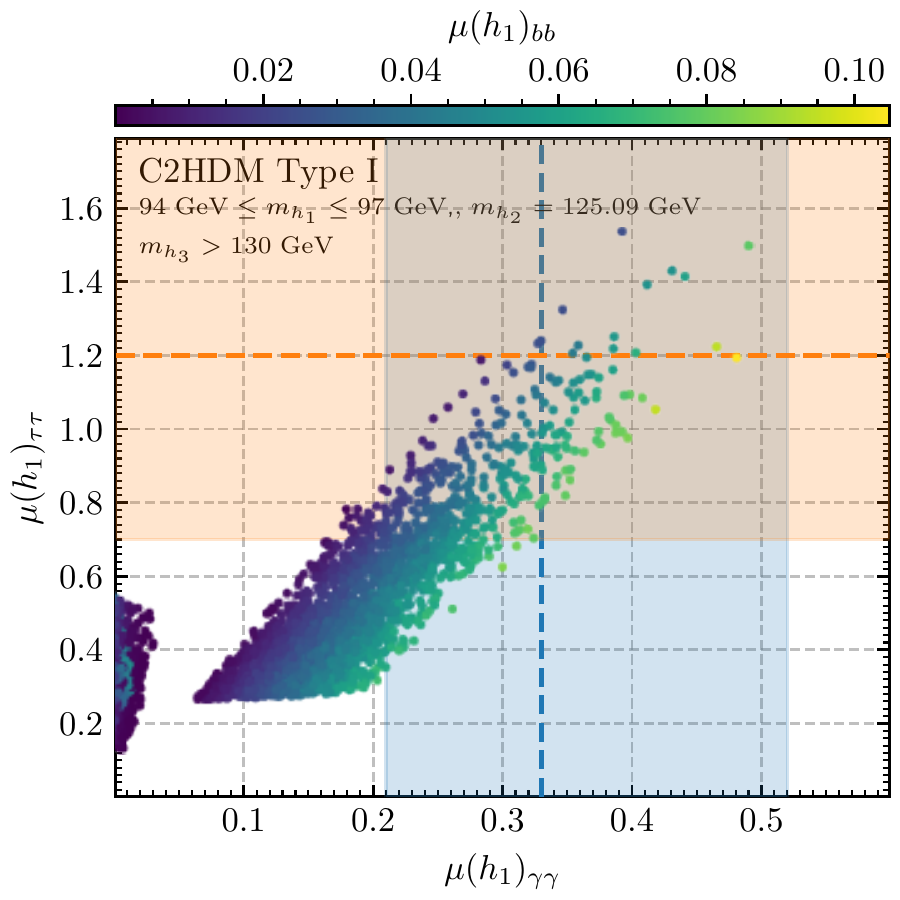}
  \caption{Diphoton signal rate
  $\mu(h_1)_{\gamma\gamma}$
  {\em vs} ditau signal rate
  $\mu(h_1)_{\tau \tau}$
  for a  95~GeV scalar within 
  the Type~I C2HDM. Except for constraints from
  flavour physics and EDMs, all
  theoretical and experimental constraints are applied.
  The colour coding of the points indicates
  the values of $\mu(h_1)_{b b}$.
  The bands correspond to the experimentally
  observed signal rates within their $1\sigma$
  uncertainty band (see \refta{tab:mus95}).}

\label{fig:mu_mu_c2hdm}
\end{figure}

In Fig.~\ref{fig:mu_mu_c2hdm}, we show how the several excesses are correlated in order to shed more light on the
possibility of describing the three excesses together.
An increase
in $\mu(h_1)_{\gamma\gamma}$ implies an increase in $\mu(h_1)_{\tau\tau}$, with predilection
for lower values of $\mu(h_1)_{bb}$. Fitting the LEP excess is relatively easy in this
version of the C2HDM, due to the larger allowed values for the coupling between $h_1$ and
$Z$ bosons compared to the real 2HDM case. Indeed,
for the real 2HDM, as seen in the previous 
section, the coupling modifier between a CP-even 95~GeV
scalar and electroweak gauge bosons
was found to be $|C_H| =
|\cos(\beta-\alpha)|\lesssim 0.2$
for the parameter points that provide a good
description of the diphoton excess.
In contrast, for the C2HDM scan with the
mass hierarchy studied in this section,
we find also values of $|C_1| \approx 0.3$ (see \refeq{eq:Cs}).

\subsubsection{Properties of $h_2 = h_{125}$}

\begin{figure}
\centering
\includegraphics[width=0.48\textwidth]{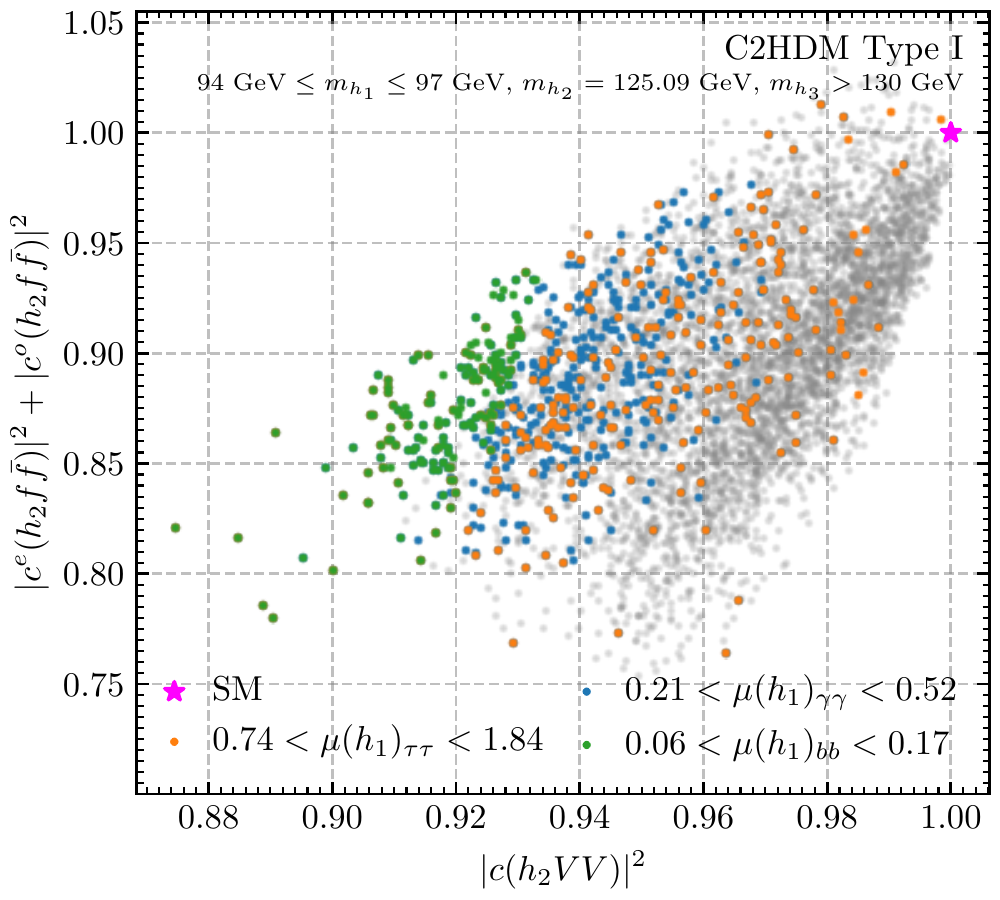}~
\includegraphics[width=0.48\textwidth]{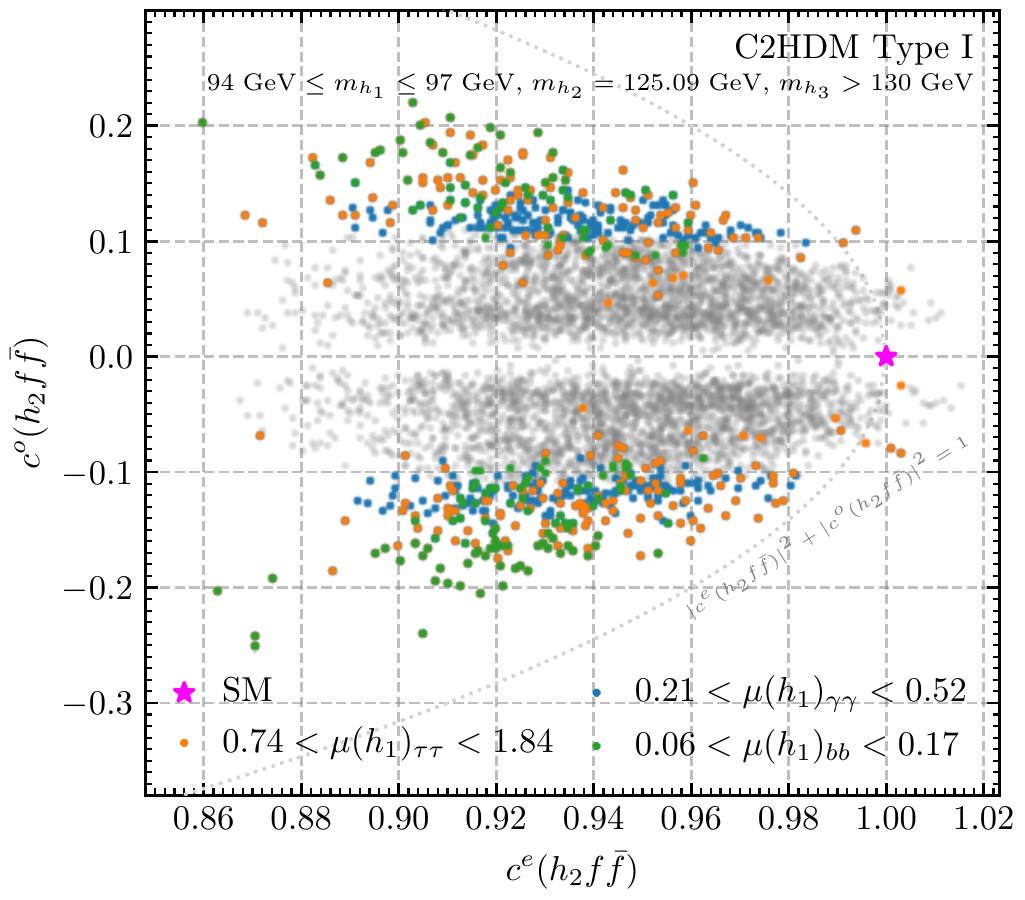}
\caption{Coupling coefficients of~$h_2 = h_{125}$ to gauge bosons
and fermions. On the left, the squared coupling of $h$ to fermions
(including both CP-even and CP-odd contributions)
vs. the respective squared coupling to gauge bosons is shown. On the right,
the CP-odd contribution to the fermion coupling of $h_2$ $c^o$
against the CP-even contribution to the same coupling, $c^e$, is shown. The dotted line
indicates a Higgs coupling to fermions with magnitude identical to the SM.
Blue, orange and green points predict a signal rate for the
  diphoton, ditau and LEP excess within
  the experimentally observed $1\sigma$
  uncertainty bands, respectively (see Tab.~\ref{tab:mus95}).
Green points are plotted on top of blue ones, which
are themselves plotted on top of orange points.
  The remaining points are shown in grey.
The SM prediction for the couplings coefficients
is indicated with a magenta star.
}
\label{fig:h125c}
\end{figure}

Compared to the R2HDM scenario discussed in \refse{sec:r2hdm}, 
the C2HDM scenario presented here has the additional
goal of accommodating the excess observed at LEP, for which
the state at~95~GeV has to be produced in Higgsstrahlung. Hence, in this case, departures from the alignment
limit of the 2HDM are required, since otherwise only
the scalar playing the role of the SM-like Higgs boson
at~125~GeV would couple to gauge bosons, while the other
neutral scalars would not couple to $Z$ bosons and accordingly
would not be produced.
Moreover, if CP-violation is taken into account,
all neutral scalar states can potentially mix with each other.
In contrast, in the CP-conserving R2HDM there is
a purely CP-odd scalar state and only the two CP-even
scalar states can mix among each other.
Given these additional sources of modifications of the
couplings of $h_2 = h_{125}$, it is even more interesting
to scrutinize the predictions for the couplings of~$h_{125}$
and the possibility of using the detected Higgs boson
at~125~GeV as a probe of the proposed C2HDM scenario.
In particular, assuming that the couplings of the state at~95~GeV
are not measured precisely, probing the couplings of the
SM-like Higgs boson might provide vital information about
the possibility of distinguishing between different models
that can describe the excesses.

In the left plot of \reffi{fig:h125c}, we show the squared coupling
coefficients, i.e.~$\kappa$-factors, for the couplings of
$h_{125}$ to gauge bosons on the horizontal axis and
for the couplings to fermions on the vertical axis, where
in the latter case we depict the squared sum of both the
CP-even and the CP-odd component of the couplings.
The predictions for the couplings of a SM Higgs boson
are indicated with the magenta star,
and the parameter points that fit the $b \bar b$-excess,
the diphoton-excess or the ditau-excess are indicated
with the green, blue and orange points, respectively.
According to the discussion above, demanding a description
of the $b \bar b$-excess observed at LEP gives rise to the
largest deviations from the alignment limit and the SM
predictions, with values of $|c(h_{125}VV)|^2 \lesssim 0.95$.
For the parameter points fitting the diphoton
excess, we find deviations from the SM of only a few
percent in the squared coupling coefficients, and the ditau-excess
can be described with deviations of only one percent.\footnote{We
note that the implementation of the C2HDM in the code
\texttt{ScannerS} does not allow for continuously approaching the
CP-conserving limit of the model. Therefore, all parameter points
of our scan in the C2HDM
will feature some amount of CP-violation.
This is why we performed a dedicated scan in the R2HDM
whose results are discussed in \refse{sec:r2hdm}.}
These deviations are smaller than the expected
experimental sensitivity with which the couplings of~$h_{125}$
can be measured during the high-luminosity runs
of the LHC. Thus, the expected precision of a future
$e^+e^-$ collider like the International Linear
Collider~(ILC) could significantly improve the prospects
of probing the 2HDM interpretation of the excesses
at~95~GeV, determining the values of the
underlying model parameters,
and also to distinguish it from other scenarios discussed
in the literature.

In order to quantify the amount of CP-violation in
the couplings of the detected Higgs boson at~125~GeV
predicted by the parameter points describing the excesses,
we depict in the right plot of \reffi{fig:h125c}
the CP-even components $c^e(h_{125} f \bar f)$
and the CP-odd components
$c^o(h_{125} f \bar f)$ of the couplings
to fermions on the horizontal and the vertical
axis, respectively. The color coding of the points is
the same as the one in the left plot of \reffi{fig:h125c}
discussed above.
One can see that a description of the three excesses
is possible with values of the CP-odd coupling
component at the level of $c^o(h_{125} f \bar f) \approx 0.1$,
corresponding to an effective CP mixing angle
of $\phi_{\rm CP} = \tan^{-1} c^o(h_{125} f \bar f)
/ c^e(h_{125} f \bar f) \lesssim 12^{\circ}$,
which is substantially below
the current experimental sensitivity at the
LHC~\cite{CMS:2021sdq,ATLAS:2022akr}.
However, under certain assumptions, the CP-violating
phases of the scalar potential of the
C2HDM can also be probed indirectly
by experimental measurements of electric dipole moments,
which will be discussed in more detail in
\refse{sec:edms}.

\subsubsection{Properties of the other BSM states}

\begin{figure}[t]
\centering
\includegraphics[width=0.5\textwidth]{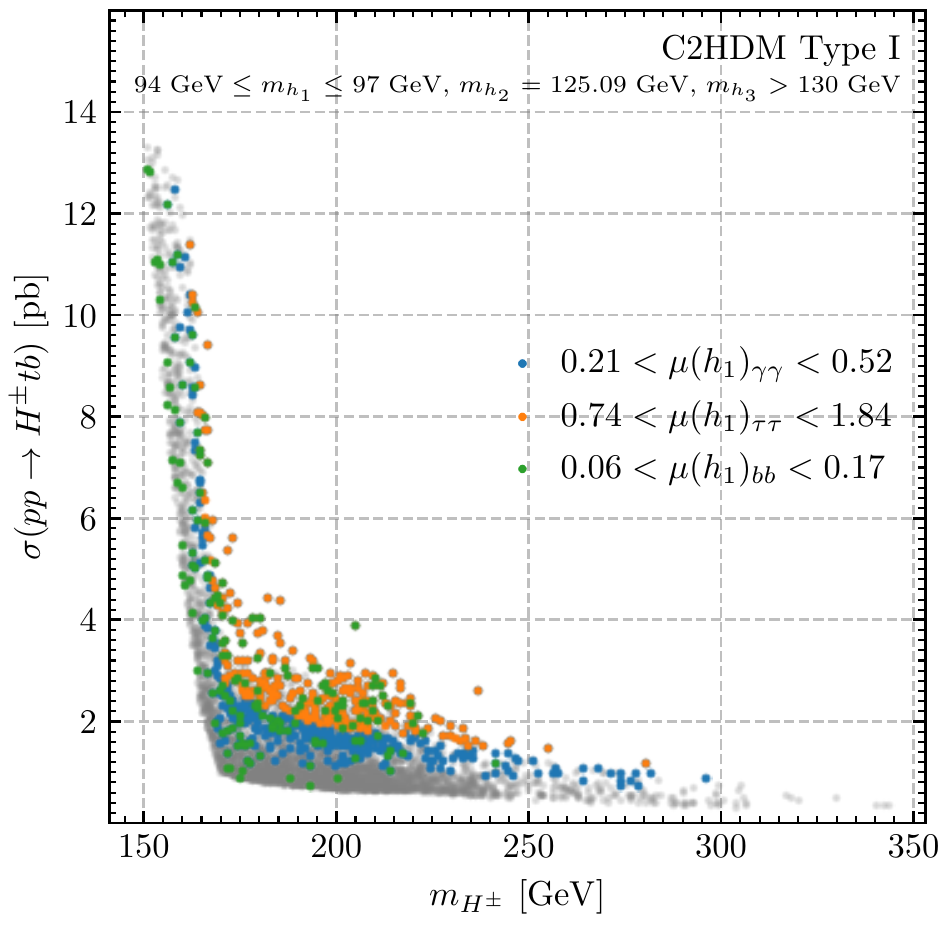}~
\includegraphics[width=0.5\textwidth]{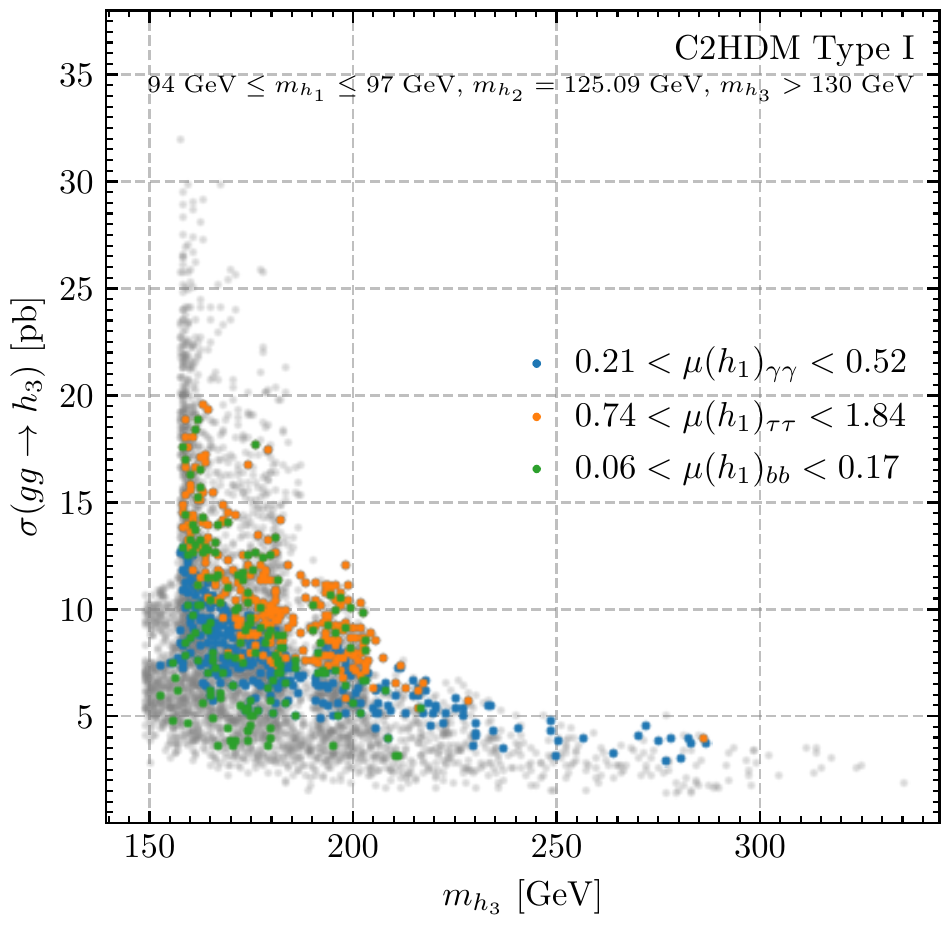}
  \caption{Allowed cross sections for $h_3$ and $H^\pm$
  production \textit{vs} 
  the respective masses. The blue, orange and
  green points describe the CMS diphoton excess,
  the CMS ditau excess or
  the LEP $b \bar b$ excess at the level of
  $1\sigma$ or better. The remaining points
  are shown in grey.
  All points satisfy the
  theoretical and experimental constraints considered previously,
  except for indirect constraints from flavour-physics
  observables and EDMs, which have not been applied here.
  }
\label{fig:sigs}
\end{figure}

In Fig.~\ref{fig:sigs}, we show the allowed masses for $h_3$ and $H^\pm$ in the region of
parameter space satisfying all of the constraints mentioned, as well as the cross sections
for their main production mechanisms.
We observe that both masses lie in the rough range between 
140~GeV and 340~GeV, with cross sections between
about 2.5~pb and 9~pb for gluon-gluon production of $h_3$,
and between 1~pb and 13~pb for top-bottom-associated production of the charged scalar.
Overall, the prospects of probing the C2HDM scenario
via searches for the heavier scalar states are very
similar to the prospects of the R2HDM scenario discussed
in \refse{sec:others}.
Future LHC searches for $H^\pm \to t b$ and $h_3 \to VV$
could potentially exclude the still viable mass interval
of $H^\pm$ and $h_3$ in the scenario under investigation.
Again, we emphasize that results for these searches
utilizing the full Run~2 dataset are so far only published
for masses above 200~GeV. One overarching message
we wish to convey,
independently of the fate of the excesses, is the
importance of covering also lower mass ranges in the
experimental searches for new Higgs bosons, if possible.

\subsubsection{Flavour-physics constraints}

\begin{figure}[t]
\centering
\includegraphics[height=8cm,angle=0]{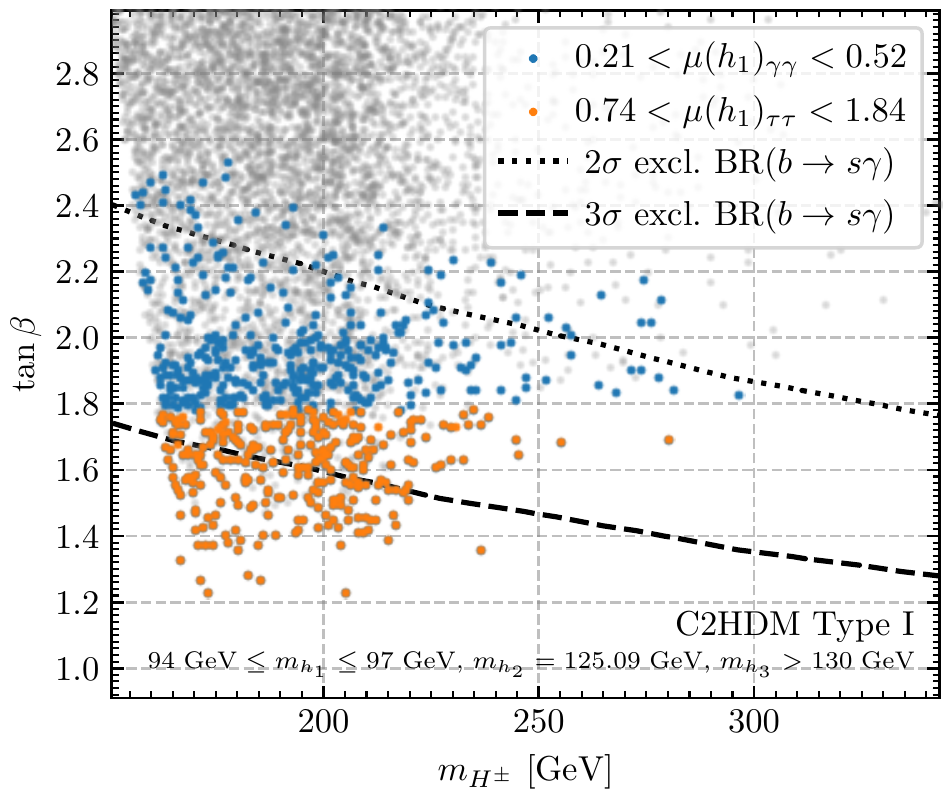}
\caption{Flavour constraints in the $\tan\beta$--$m_{H^\pm}$ plane.
Blue and orange points indicate the parameter points that
describe the diphoton and the ditau excess
at the level of $1\sigma$ or better, respectively,
where orange points are plotted on top of blue points.
The remaining parameter points are shown in grey.
Parameter points fitting the LEP excess are not indicated
here, since the corresponding signal rates are not
strongly correlated with the values of $\tan\beta$
(see \reffi{fig:tbeta_mus_c2hdm}).
The dotted and the dashed black lines indicate the
exclusion limit from the measurements of $b \to s \gamma$
branching ratios at the $2\sigma$ and the $3\sigma$ confidence
level, respectively, where the excluded regions are the parameter
region below these lines. }
\label{fig:fl_C2HDM}
\end{figure}

As in the R2HDM, the LHC diphoton and
ditau can be accomodated within the Type-I C2HDM,
with the added bonus that the LEP bottom excess can
be fitted as well. Unlike the R2HDM, however,
the region where two of these excesses are accommodated by
the C2HDM is {\em not} necessarily in significant
tension with flavour physics constraints.
This much may be seen from \reffi{fig:fl_C2HDM},
wherein we plot (in grey)
all points from our scan in the $\tan\beta$--$m_{H^\pm}$ plane. 
Blue points denote the subset of points for which
the LHC diphoton excess at 95 GeV are
accommodated.\footnote{The region where the LEP
excess is described as well would not
differ much from the region in which the blue points
are located in \reffi{fig:fl_C2HDM}, such that
we do not highlight these points in the plot
separately.}
We see that, unlike the case of the R2HDM,
higher values of $\tan\beta$ are
compatible with a description of the diphoton excess,
and a sizeable number of points lie above the
black dotted line and are thus allowed by 
the most stringent flavour physics constraints
considered in \citere{Arbey:2017gmh}, to wit limits on 
the $b\to s \gamma$ branching ratio. If one
further attempts to describe the LHC ditau excess
at 95~GeV, one is the restricted to the parameter
points shown in orange, which lie below the
$2\sigma$ confidence-level exclusion
line from flavour physics --
this is due to the lower values
of $\tan\beta$ required to accommodate
this third signal, as had already been
discussed in Fig.~\ref{fig:tbeta_mus_c2hdm}.
However, a subset of orange points is above the
dashed black line indicating the
$3\sigma$ confidence-level exclusion limit
from $b \to s \gamma$ measurements.

\subsubsection{EDM constraints}
\label{sec:edms}

Given the new source of CP-violation in
the scalar potential of the C2HDM,
another way of probing the parameter points
fitting the excesses is via potentially
sizeable contributions to the static electric
dipole moment~(EDM) of fundamental particles.
In particular for the electron,
very precise measurements compatible with a
vanishing EDM typically impose the
strongest constraints on the parameter space of
the C2HDM~\cite{Fontes:2017zfn}.
In order to
confront the parameter space suitable for
a description of the excesses at~95~GeV
with these constraints,
we use the prescription from \citere{Abe:2013qla}
to compute the electron EDM and compare
the theoretical predictions with the latest
experimental upper limits at 90\%
confidence level published by the
JILA Collaboration~\cite{Roussy:2022cmp},
\begin{equation}
|d_e|<4.1\times 10^{-30}e~\textrm{cm} \ ,
\end{equation}
where $e$ is the electric charge of the
electron. 
In our parameter scan of the C2HDM,
the maximum values obtained are of the
order of $|d_e| \approx \mathcal{O}(10^{-27})$,
with the bulk of points predicting an
electron EDM at the level of
$|d_e| \approx \mathcal{O}(10^{-28})$.
Consequently, strictly
applying the experimental upper limit from
JILA would exclude all parameter points of our scan.

However, we stress that there are sizable theoretical
uncertainties in the applications of these limits:  the theory predictions
are sensitive to the specifics of the UV completion of the 2HDM~\cite{Jung:2013hka},
even if the additional new physics lives at multi-TeV energy scales~\cite{Cesarotti:2018huy}.
Moreover, focusing on the 2HDM particle content, large cancellations can be present between different
contributions to the EDMs from the different scalars (for instance, if almost mass degenerate neutral
scalar states are present~\cite{Fontes:2014xva}), and also additional sources of CP-violation in the
Yukawa sector (not considered here) can play a role
for the theory predictions of the EDMs (see e.g.~\citere{Fuyuto:2019svr} for a
discussion in the general 2HDM). Since we want to remain agnostic about the UV
completion of the 2HDM, and since we focus here on CP-violating effects only from the Higgs sector,
whereas the EDMs also depend on the CP-phases in the fermion sector, we do
not attempt a dedicated analysis with the goal of suppressing the electron EDM in order to be
compatible with the limit from the JILA collaboration.

\subsection{Complex 2HDM -- $m_{h_1} = 95$~GeV, $m_{h_2} = 125$~GeV, $m_{h_3} < 120$~GeV}
\label{sec:mli}

In the previous sections we considered the possibility of $h_{125} = h_2$ being the
the second lightest neutral scalar. However, having a scalar $m_{h_3} < 120$ GeV is also
allowed under the different
theoretical and experimental cuts, in which case the detected Higgs boson at~125~GeV
corresponds to the heaviest neutral scalar state. This mass hierarchy would imply that two neutral scalars
below 125~GeV would have escaped detection so far, and we present those scenarios here for completeness. The 
results from our scans are shown in
\begin{figure}[t]
\centering
\includegraphics[width=0.33\textwidth]{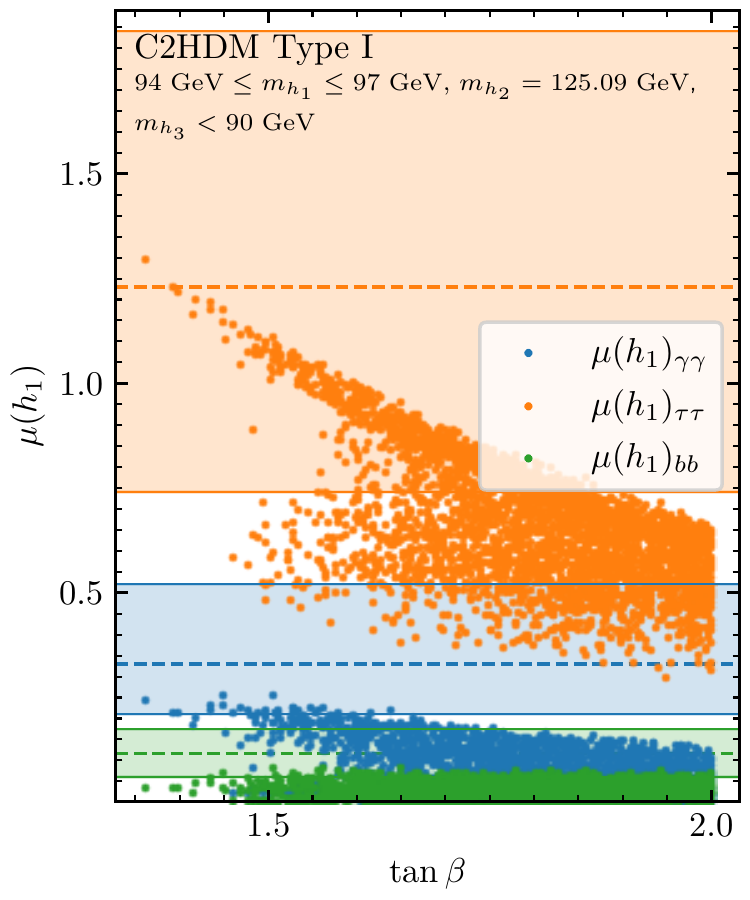}~
\includegraphics[width=0.33\textwidth]{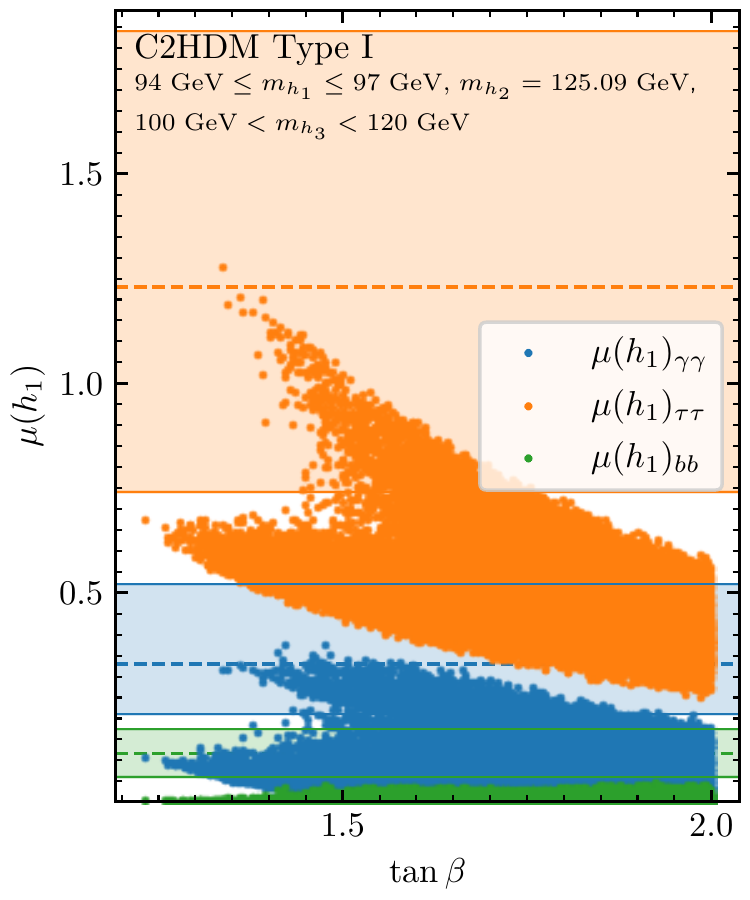}
  \caption{Signal rates for the diphoton,
  ditau and $b \bar b$ excesses {\em vs}
  $\tan\beta$ for a
  95~GeV scalar within 
  the Type~I C2HDM and for different
  mass hierarchies. Except for constraints from
  flavour physics and EDMs, all
  theoretical and experimental constraints are applied.
  The bands correspond to the experimentally
  observed signal rates within their $1\sigma$
  uncertainty band (see Tab.~\ref{tab:mus95}).
  }
\label{fig:tbeta_mus_c2hdm_mli}
\end{figure}
Fig.~\ref{fig:tbeta_mus_c2hdm_mli}, for the two ranges 
$m_{h_3} < 90\gev$ and $100\gev < m_{h_3} < 120\gev$, arising from the
condition of not having mass degenerate states. The left plot
shows that if $m_{h_3}  < 90\gev$,
and the state $h_1$ at about~95~GeV
is the second lightest of the three neutral scalars, a fit of the three signals -- diphoton,
ditau, LEP $b\bar{b}$ -- is barely possible for $1.0\lesssim \tan\beta\lesssim 1.7$.
However, notice that the fit to the diphoton excess is 
very difficult to achieve in this scenario, with predicted signal rates
at the lower end of the $1\sigma$ region. The plot on the right shows that,
for the intermediate mass hierarchy, $100\leq m_{h_3} \leq 120$ GeV, 
fitting the LEP signal is not possible, but the
diphoton and ditau excesses may be fitted for $1.35\lesssim \tan\beta\lesssim 1.75$.

\begin{figure}
\centering
\includegraphics[width=0.5\textwidth]{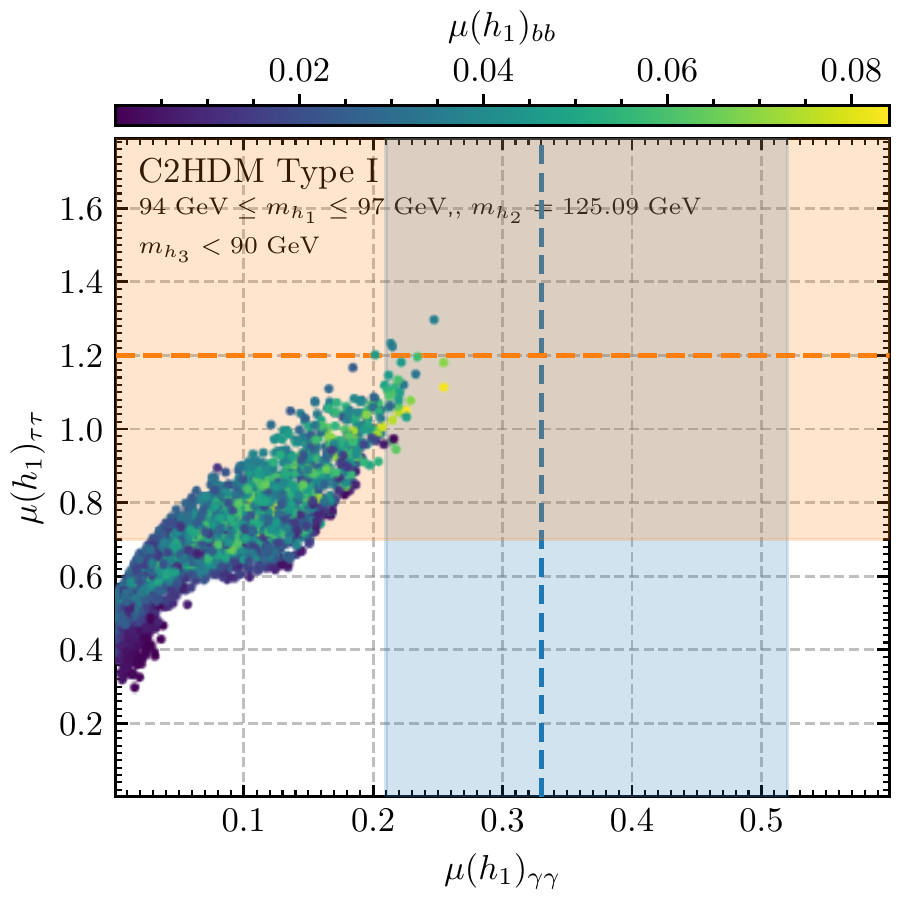}~
\includegraphics[width=0.5\textwidth]{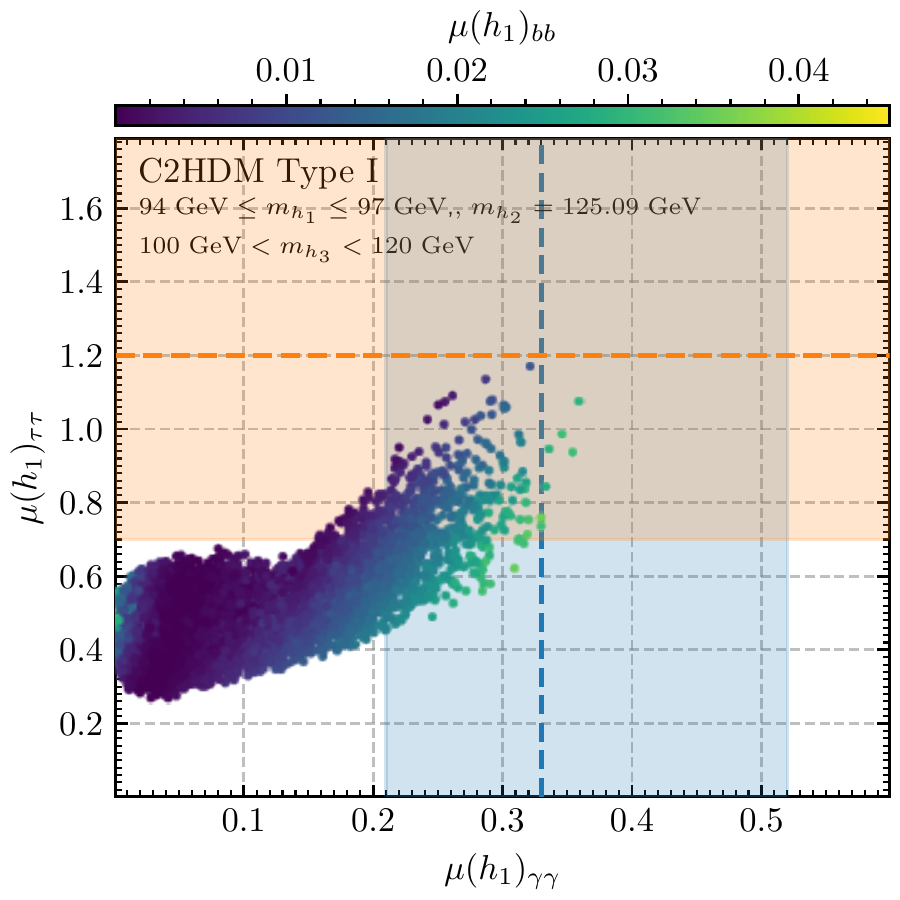}
  \caption{Diphoton signal rate {\em vs} ditau signal rate
for a  95~GeV scalar within 
  the Type~I C2HDM. Except for constraints from
  flavour physics and EDMs, all
  theoretical and experimental constraints are applied.
  The bands correspond to the experimentally
  observed signal rates within their $1\sigma$
  uncertainty band (see Tab.~\ref{tab:mus95}).
  }
\label{fig:mu_mu_c2hdm_mli_h3low}
\end{figure}

In \reffi{fig:mu_mu_c2hdm_mli_h3low}
we again show the correlation between the several $\mu$-values, where it is easy to appreciate that the
a sufficiently large diphoton signal is
very difficult to accomplish for $m_{h_3} \leq 90$~GeV;
and that the LEP fit cannot be accomplished if $100\gev \leq m_{h_3} \leq 120$~GeV.
These light-$h_3$ scenarios are, therefore, 
less interesting than the situation considered in
the previous sections, where $h_3$ is the heavier of the neutral scalars. 
For completeness, as with the first mass hierarchy studied, the points fulfilling both the ditau and diphoton excesses are at most between the $2\sigma$ and $3\sigma$ line of the $b\to s\gamma$ measurement. 

\subsection{Complex 2HDM -- Lepton-specific}

As a final note, we also investigated the Lepton-specific version of the C2HDM. Therein, the generated points are within the region $\tan\beta<1$ and $m_{H^\pm}<200$ GeV, such that points that can fulfill the signal ranges for the diphoton and ditau final states are in strong tension with flavour constraints at more than $3\sigma$ (\textit{c.f.} Fig.~\ref{fig:fl_C2HDM}). Therefore, we consider this scenario as 
excluded by current experimental data and will not discuss
it in detail here.

\section{Conclusions}
\label{sec:conc}

The idea that there might be lighter scalars than the Higgs boson 
discovered at LHC in 2012 is quite enticing and it is not at all excluded by current
experimental measurements. On the contrary, as the LHC accumulates data and searches for
BSM particles become more refined, several interesting possibilities arose. Searches
for diphoton decays at CMS have, since 2018, consistently indicated a (less than $3\sigma$) 
excess for an invariant mass of 95 GeV. A possible ditau excess at a compatible mass is also 
observed by CMS, and, of course, that mass value reminds us of the presumed $b\bar{b}$ excess 
seemingly observed in the final days of LEP.

To interpret these possible signals as a spin-0 particle
it would be desirable to have a ``natural" mechanism to explain why a lighter scalar 
would only now be discovered, after the heavier SM-like 125 GeV Higgs boson discovery.
The 2HDM provides a simple explanation for that -- a CP-even 95 GeV scalar would have its 
couplings to electroweak gauge bosons suppressed due to the SM-like behaviour of the 125 GeV
scalar; and a pseudoscalar of the same mass would not have been observed at LEP (thus
the LEP anomaly would have been simply an experimental fluctuation) but could provide a nice explanation
to the current LHC excesses at 95 GeV. Flavour physics then dictates that, out of the different 
flavour-preserving versions of the 2HDM, models Type II or Flipped are ruled out, since $b\rightarrow s\gamma$
bounds impose a charged mass of several hundreds of GeV, and it becomes impossible to accomodate (due to 
unitarity constraints, as well as electroweak precision bounds) a lighter, 95 GeV, neutral state.

Our analysis shows that within the framework of the CP-conserving Type I 2HDM (R2HDM), a CP-even scalar with
mass of 95 GeV could not fit both the ditau and diphoton data, unless very small values of $\tan\beta$ -- highly
disfavoured by flavour constraints -- would be taken into account. On the contrary, assuming a pseudoscalar with
mass 95 GeV, the higher production cross sections for such a particle (compared with a  CP-even scalar of the same mass)
allow for a simultaneous fit of both diphoton and ditau signals, though the LEP excess could not be fitted in
this scenario. Unlike other 
attempts to fit these signals
in the framework of the 2HDM, within the
scenario proposed here, the only production 
mechanism required is gluon-gluon fusion. The resulting masses for the heavier CP-even scalar $H$ and the charged Higgs $H^\pm$ are relatively light (between roughly 160 and 250 GeV), and production cross sections for 
these BSM states are predicted by the model to range from 2 to 15 pb. The model will be, therefore, 
probed
at Run~3 of the LHC, and we have shown that (again unlike other attempted theoretical explanations for these results)
a fit of both diphoton and ditau data is possible even with minimal deviations from the alignment limit. There is,
however, a problem with this scenario -- $b\rightarrow s\gamma$ constraints disfavour, at 2.5$\sigma$ or more, the range of
charged masses and $\tan\beta$ required to fit both LHC excesses. 

We were, in turn, led to the C2HDM, where CP violation in the potential is introduced via a soft $\mathbb{Z}_2$-breaking term. The model
has more freedom than the R2HDM to
be in agreement with
current experimental constraints. Within this more general framework, we find 
a fit to the three excesses -- LHC diphoton and ditau, LEP $b\bar{b}$ -- is possible for larger values of $\tan\beta$.
Indeed, it even becomes possible to fit the diphoton and LEP signals (not the ditau one) so that the tension with
$b\rightarrow s\gamma$ is now inferior to 2$\sigma$. With CP violation induced by the scalar potential, however, other bounds,
such as those arising from electron EDMs, need to be considered. A quick analysis 
indicates that the region of parameter space 
suitable for a good description of
all three signals would be in conflict with electron EDM data. Once again, the fit prefers masses for the charged and third neutral
scalar below roughly 250 GeV and predicts sizeable production
cross sections for those particles at the LHC.

Recently, the ATLAS collaboration has reported
the full Run~2 results for searches for
low-mass Higgs bosons
in the di-photon final state. The model-dependent ATLAS
analysis has a very similar sensitivity
to the~CMS search, showing the most significant excess
over the~SM background  at a mass of about
95~GeV, consistent  with the CMS di-photon excess.
However, the excess observed by~ATLAS
is much less pronounced, such that
a combination between CMS and ATLAS would yield
slightly smaller signal rates for a possible state
at~95GeV.
We have checked that combining the latest ATLAS diphoton search 
would not change the previous statements
qualitatively: the three excesses can still be described
simultaneously, although potentially with slightly larger
$\tan\beta$-values.

The remaining allowed mass interval
for the charged Higgs bosons
is found in the vicinity of the top-quark mass, where
the LHC searches
for $H^\pm \to tb$ lose sensitivity, but for which the
exotic top-quark decay $t \to H^\pm b$ is still
kinematically suppressed.
In a similar fashion, assuming the mass hierarchy
with $h_{125}$ being the second lightest scalar state,
the remaining allowed mass window
for the third neutral scalar is found in the neighbourhood
of twice the $Z$-boson mass.
Larger values of $m_{h_3} \gtrsim 250\gev$
give rise to too large cross sections for the
signature $pp \to h_3 \to h_1 Z$ with $h_1 = h_{95}$,
which was searched
for at the LHC assuming that $h_1$ decays in
tau-lepton or bottom-quark pairs.
On the other hand, values of $m_{h_3} \lesssim 160\gev$
are in tension with cross-section limits from LHC searches
for $h_3 \to \tau^+ \tau^-$.
An overarching result of our study is that,
independently of the fate of the excesses, 
LHC searches for rather light Higgs bosons
at or around the EW scale still have significant
potential to probe so far allowed parameter
space regions of the 2HDM, in particular
searches for Higgs bosons decaying into gauge bosons
with masses below 200~GeV.

Further analysis of these possibilities for the Lepton-specific model, and for mass hierarchies with two scalars lighter
than 120 GeV, reveals some interesting possibilities, but added tension with flavour constraints.

Clearly the preferred
charged scalar mass and $\tan\beta$ region necessary for fitting the three excesses has tension with existing flavour sector results.
Here it should be noted that these tensions
would be somewhat weaker assuming a slightly smaller
diphoton signal rate as suggested by the recent
ATLAS search utilizing the full Run~2 dataset.
We argue, though, that the tensions with data from flavour physics
are not sufficient to invalidate this proposed solution for the LHC and LEP excesses at masses
about 95 GeV. If the LHC excesses are confirmed 
by Run~3 data and still explainable by the 2HDM
scenarios herein proposed, and tension with B-physics results persist, that could indicate, rather than the exclusion of the 2HDM, 
the need for a more complex flavour sector, with hitherto undiscovered BSM physics.  

In short, we proposed several very economic theoretical frameworks for reproducing the 
CMS diphoton and ditau and the LEP $b\bar{b}$
excesses -- so economical that, within the context of the R2HDM with a pseudoscalar of mass 95 GeV, the analysis
can almost be made analytically.
Confirmation of the excesses observed
at the~LHC by the Run~3 measurements is paramount.
Reduced signal strengths for the diphoton and ditau signals would allow for larger values of 
$\tan\beta$ and weaken 
tensions with flavour physics constraints. Regardless, the explanations for the 95 GeV results
within a 2HDM framework put forth in this work will be tested in Run~3 and will be easily disproven -- or 
confirmed.\footnote{Considering the authors' names, if the 95 GeV excesses 
discussed in this paper are experimentally confirmed and proven to be described by a 2HDM scalar,
we propose that it be named the ABF, or even better, the FABulous particle.}

\section*{Acknowledgements}
The authors
thank Georg Weiglein for bringing
\citere{ATLAS:2023azi} to our attention.
T.B.~thanks Daniel Winterbottom for
interesting discussions.
P.M.F.~is supported
by \textit{Funda\c c\~ao para a Ci\^encia e a Tecnologia} (FCT)
through contracts
UIDB/00618/2020, UIDP/00618/2020, CERN/FIS-PAR/0004/2019, CERN/FIS-PAR/0014/2019 and CERN/FIS-PAR/ 0025/2021.
The work of T.B.~is supported by the German
Bundesministerium f\"ur Bildung und Forschung (BMBF, Federal
Ministry of Education and Research) – project 05H21VKCCA. 
D.A. is supported by the Deutsche Forschungsgemeinschaft (DFG, German Research Foundation) under grant 396021762 - TRR 257.


\bibliographystyle{JHEP}
\bibliography{biblio}

\end{document}